# Title: Population-Specific Genetic and Non-Genetic Influences on Sleep Traits and Health Outcomes

**Authors:** Jiheum Park, PhD[1*], Stephanie Y. Shue, BA[1*], Rocio Barragan, PhD[1,2,3,4], Jeong Yun Yang, MD[1], Tian Gu, PhD[5], Chin Hur, MD, MPH[1†], Marie-Pierre St-Onge, PhD[1,4†#]

**Affiliations:**

[1]Division of General Medicine, Department of Medicine, Columbia University Irving Medical Center, New York, NY 10032, USA

[2]Department of Preventive Medicine and Public Health, School of Medicine, University of Valencia, 46010 Valencia, Spain.

[3]CIBER Fisiopatología de la Obesidad y Nutrición, Instituto de Salud Carlos III, 28029 Madrid, Spain.

[4]Center of Excellence for Sleep & Circadian Research, Department of Medicine, Columbia University Irving Medical Center, New York, NY 10032, USA

[5] Department of Biostatistics, Columbia University; New York, NY, USA

# Corresponding author

Marie-Pierre St-Onge, PhD
622 West 168th Street, PH9-103H
New York, NY 10032
ms2554@cumc.columbia.edu

[*] These authors contributed equally

†Co-senior authors

**Abstract**

**Purpose**: Sleep traits, shaped by both genetic and environmental factors, influence various physiological conditions. Diverse data representing the U.S. population from the All of Us (AoU) Research Program — including electronic health records (EHR), physical measurements, genomic information, and wearable device data across ancestry groups — offers a unique opportunity to explore the interplay between genetic and non-genetic factors in sleep traits and their associations with health outcomes and disparities. This study aims to examine the associations between genetic predispositions to sleep traits (chronotype, sleep duration, and short sleep) and health outcomes across ancestries, as well as the influence of actual sleep duration.

**Methods:** We leveraged AoU genome-wide association study (GWAS) results, including ancestry-specific and meta-analyses for 3,414 phenotypes, to identify phenotypes associated with 455 sleep-related SNPs. Cross-sectional and longitudinal analyses (n = 212,529) evaluated the associations between polygenic risk scores (PRS) for sleep traits and anthropometric/metabolic measures from EHR. A subgroup analysis (n = 7,655) assessed the influence of objectively measured sleep duration using Fitbit data.

**Results**: SNP analysis across six ancestry groups identified 61 phenotypes linked to 29 sleep-trait-associated SNPs. The chronotype SNP rs1421085 in the fat mass gene showed the strongest associations with anthropometric, obesity, diabetes, and cardiovascular conditions in the meta-analysis. These associations were primarily observed in European, American, and African groups in ancestry-specific analyses. PRS analysis indicated that a higher predisposition to shorter sleep duration was linked to increased risk of both obesity and diabetes, however with ancestry-specific variations. Objectively measured sleep duration acted as a confounder, rendering these associations non-significant, with relative contributions ranging from 85.6%–99.9% (cross-sectional) and 7.1%–44.0% (longitudinal) compared to PRS.

**Conclusion**: This study identified health conditions associated with genetic predispositions to sleep traits, with implications that actual sleep duration may play a more prominent role in sleep-related health outcomes. Differences among meta-, pooled-, and ancestry-specific analyses underscore the importance of population-specific research.

**Introduction**

Sleep is an essential state that affects nearly every system in the body, influencing a wide range of physiological and cognitive processes. Sleep traits impact various health outcomes, including cardiovascular[1], metabolic[2], immune[3], and cognitive[4] functions. Chronotype, an individual's preference for earlier or later activities, has been linked to an increased risk of developing metabolic[2] and psychiatric[5] disorders. Similarly, both short (<6h) and long (>9h) sleep durations have been associated with cognitive, psychiatric, metabolic, and cardiovascular dysfunction.[6] However, directly measuring individuals' sleep traits and their associations with health outcomes requires extensive long-term clinical studies. Genetic data offer an effective alternative for estimating sleep traits and their relationships with various health outcomes, potentially leading to the development of effective interventions.

With the advent of Genome-Wide Association Studies (GWAS), a multitude of gene variants, most commonly Single Nucleotide Polymorphisms (SNPs), have been linked to complex diseases and traits, including cardiovascular disease, cancers, obesity, autoimmune diseases, and more.[7] By summing the weighted effect of SNPs identified in GWAS for a specific trait, an estimate of an individual's genetic risk for that trait can be generated. This estimate, known as a Polygenic Risk Score (PRS), can be a powerful tool for understanding how genetic variations contribute to various health outcomes, offering potential pathways for personalized medicine and targeted therapies.

Recent GWAS have identified genetic variants associated with chronotype, short sleep, and sleep duration.[8,9] In a cohort of 697,828 UK Biobank and 23andMe participants, predominantly of European ancestry, 351 SNPs have been found to be associated with chronotype.[8] Additionally, in a cohort of 446,118 adults of European ancestry from UK Biobank, 78 SNPs have been associated with sleep duration, including 27 SNPs specifically linked to short sleep (<7h per night).[9]

These genetic risk markers for sleep traits, identified through GWAS, may reveal meaningful relationships with health outcomes, potentially informing effective interventions. However, numerous studies have demonstrated the significant influence of environmental factors, including social conditions (e.g., neighborhood disorder) and physical features (e.g., light, noise), on sleep patterns.[10] A meta-analysis on heritability of sleep duration and quality found that 46% of the variability in sleep duration and 44% of the variability in sleep quality are genetically determined, with remaining variation largely attributed to unique environmental factors experienced by individuals. Shared environmental influences also appear to play a role in childhood sleep duration, with heritability estimates varying significantly by age.[11]

In this study, we aim to investigate the associations between genetic and non-genetic influences on sleep traits and health outcomes using diverse datasets from the All of Us (AoU) Research Program, which include electronic health records (EHR), physical measurements, genomic information, and wearable device data across different ancestry groups.[12]

We hypothesized that leveraging this representative, large-scale, individual-level longitudinal health data, coupled with genomic information across diverse ancestries in the AoU dataset, would provide valuable insights into the genetic underpinnings of sleep health and its relationship to various health conditions. Additionally, the availability of actual sleep measurements from Fitbit data for a subset of participants offers a unique opportunity to explore the interplay between genetic predispositions and non-genetic factors in shaping sleep-related health outcomes.

## Methods

All the analyses were conducted using Python 3 in the AoU Researcher Workbench, and the overall study workflow is shown in Figure 1. We utilized five datasets from AoU Research Program – Genomic data, one-time physical measures data, longitudinal EHR data, Fitbit sleep data, and All by All tables which include association test results across six ancestry groups (African (AFR), Admixed American (AMR), European (EUR), East Asian (EAS), South Asian (SAS), and Middle Eastern (MID)) and a meta-analysis. Detailed descriptions of each dataset, data processing with inclusion/exclusion criteria[13-15] (eFigure 1), and the AoU enrollment[12,16,17] are provided in the Supplementary Material.

**Sleep-trait-associated SNPs.** We obtained SNPs linked to three sleep traits – chronotype (351 SNPs), short sleep (27 SNPs), and sleep duration (78 SNPs) – from published GWAS.[8,9] We identified 446 SNPs in All by All tables and 451 SNPs (346 for chronotype, 27 for short sleep and 78 for sleep duration) in AoU genomic dataset.

**Association test for individual sleep-trait-associated SNPs with phenotypes.** We used ancestry-specific and meta-analysis results from All by All tables to discover phenotypes significantly associated with the identified sleep-trait-associated SNPs. Significance was determined using a p-value threshold <5x1e-8 to limit false positives.[18] SNPs in the meta-analysis table were further filtered by a p-value for heterogeneity threshold of >0.05 to identify phenotypes that were significant across ancestries.

**PRS for sleep trait and association test with phenotypes.** We calculated three separate PRS for each participant included in the analysis (Table 1): PRS-C, PRS-SS, and PRS-SD, where C, SS, and SD indicate chronotype, short sleep, and sleep duration, respectively. We used the AoUPRS package[19] for the PRS calculation. Additional details on the PRS calculation are provided in the Supplementary Material.

*Cross-sectional analysis:* We performed multivariable linear regression for each pair of sleep trait PRS and measurement, with the formula provided in the Supplementary Material. For each participant, the median value and age at measurement were obtained from the longitudinal BMI, waist circumference (WC), hemoglobin A1C (HbA1C), and fasting glucose and insulin measurements. We adjusted for demographic covariates including sex at birth, genetic ancestry group, ethnicity, and age, as well as additional covariates such as history of smoking, obesity, and diabetes (eTable 1). The median age at measurement was used for the age covariate. Demographic data in AoU were pulled from EHR where race, sex, and ethnicity are self-reported. As a considerable portion of our cohorts did not indicate race (17.2%), we opted to use genetic ancestry as a surrogate for race. Predicted genetic ancestry is available for all participants with srWGS in AoU.

*Longitudinal analysis:* We employed linear mixed models to account for the correlation between longitudinal measurements within individuals while assessing the association between each sleep trait PRS and measurement pair. For each measurement type, we excluded values from participants who had less than three measurements. No participant had more than two WC measurements, so WC was excluded from the longitudinal analyses. Further details on the use of linear mixed models, including the formula, are available in the Supplementary Material.

*PRS vs sleep duration contribution analysis:* For both cross-sectional and longitudinal analyses, we assessed associations after adjusting for average daily sleep duration derived from Fitbit sleep data. To confirm findings, we also evaluated the association between sleep duration and health outcomes with adjustments for PRS. To explore potential causal relationships, mediation analysis was performed using the built-in Python package from statsmodels. The relative contributions of PRS and actual sleep duration to health outcomes were quantified using coefficients from linear mixed models. Since the data were standardized, these contributions were calculated by comparing the squared coefficients of PRS and sleep duration.

## Ethics Statement

Ethical approval for this study was granted by the All of Us Institutional Review Board (IRB). The All of Us Institutional Review Board follows the regulations and guidance of the National Institutes of Health Office for Human Research Protections, ensuring the rights and welfare of research participants are consistently overseen and protected. All participants in the All of Us Research Program provided written informed consent before data collection. Data privacy and security were maintained per AoU policies, ensuring compliance with regulatory requirements.

## Results

### Phenotypes associated with individual sleep-trait-associated SNPs

In the meta-analysis, we identified 23 unique phenotypes significantly associated with 11 sleep-trait-associated SNPs (8 for chronotype, 1 for sleep duration, and 2 for short sleep) (Figure 2). Significant associations were found across ancestry groups in EUR, AMR, and AFR, with 61 unique phenotypes linked to 29 sleep-trait-associated SNPs (18 for chronotype, 8 for sleep duration, and 3 for short sleep) (Figure 3). These identified phenotypes were grouped into 8 health categories: anthropometric measures, obesity, diabetes, cardiovascular disease, lipid metabolism, sleep disorders, neurological disorders, and blood tests (eTable 2).

In both meta-analysis and ancestry-specific analyses, SNPs associated with chronotype showed stronger associations and larger effect sizes with phenotypes compared to those linked to sleep duration and short sleep (Figures 2a and 3 for -log(p-value) and Figure 2b and eFigure 2 for effect sizes). Notably, the chronotype SNP rs1421085 (chr16:53767042), located in the fat mass and obesity-associated (FTO) gene region[20] showed the strongest association across various phenotypes, including anthropometric measures, obesity, diabetes, and cardiovascular disease.

However, we also observed discrepancies between the meta-analysis and ancestry-specific analysis. For example, the short sleep SNP rs2820313 showed significant associations with diabetes and neurological disorders in the meta-analysis but not in ancestry-specific analyses. Conversely, the sleep duration SNP rs9940646 (chr16:53766717) was not significant in the meta-analysis but showed strong associations with phenotypes related to diabetes, obesity, and cardiovascular disease in EUR group. Similarly, in EUR group, the chronotype SNP rs113851554 (chr2:66523432) was associated with phenotypes related to sleep and neurological disorders, which was not observed in the meta-analysis. Anthropometric phenotypes such as height, BMI, weight, WC, hip circumference (HC), weight-to-height ratio (WHR), and WHR adjusted BMI (WHRadjBMI) showed significant associations with all three sleep-trait-associated SNPs in the EUR, AMR, and AFR groups but not with sleep duration SNPs in the meta-analysis (Figures 2 and 3).

**Phenotypes associated with the sleep trait PRS**

To evaluate the cumulative impact of multiple genetic variants linked to each sleep trait on the phenotypes, we calculated PRS for individual participants and conducted association analyses using cross-sectional and longitudinal data from their EHRs. We specifically investigated measurements of BMI, WC, fasting glucose and insulin, and HbA1C.

*Cross-sectional analysis*

In our pooled cross-sectional analysis, PRS for all three sleep traits were significantly associated with at least one anthropometric measure (Table 2). Our results suggest that genetic risk for shorter sleep duration is associated with higher BMI (PRS-SS: $\beta=0.012$, $p<0.001$; PRS-SD: $\beta=-0.022$, $p=0.031$) and WC (PRS-SS: $\beta=0.011$, $p<0.001$; PRS-SD: $\beta=-0.016$, $p<0.001$), whereas morning chronotype PRS is associated with higher WC but not BMI. Additionally, PRS-SD was significantly negatively associated with insulin levels ($\beta=-0.070$, $p=0.031$), a marker of diabetes risk.

In the ancestry-specific analyses (eTable 3), significant associations between PRS and anthropometric measures were observed in EUR, AFR, AMR, and EAS. The findings for EUR were entirely consistent with the pooled analysis. Although not all significant associations were replicated in other groups, those that were found maintained the same directional trend as in the pooled analysis.

Regarding glycemic measures, insulin levels were significantly associated with PRS-SD in the pooled analysis. However, no significant associations between insulin and PRS for sleep traits were observed within any specific subgroup. Of note, ancestry-specific analyses revealed significant associations absent in the pooled analysis, such as correlations between PRS-SS and HbA1C, PRS-SS and fasting glucose, and PRS-SD and fasting glucose. The direction of these associations varied across ancestry groups. For instance, while short sleep duration was associated with increased diabetes risk in AMR (PRS-SS and HbA1C: β=0.019, p-0.014) and EAS (PRS-SS and fasting glucose: β=0.160, p=0.017) populations, it was linked to decreased diabetes risk in AFR (PRS-SS and HbA1C: β=-0.017, p=0.013) and MID (PRS-SD and fasting glucose: β=0.427, p=0.013) populations. These findings suggest that a genetic predisposition to short sleep is associated with higher diabetes risk markers in some groups but lower markers in others.

*Longitudinal analysis*

In our pooled longitudinal analysis, PRS-SS (β=0.017, p<0.001) and PRS-SD (β=-0.016, p=0.001) showed significant associations with BMI over time (Table 3). Additionally, PRS-SS was significantly positively associated with insulin levels (β = 0.134, p = 0.037), indicating that a genetic predisposition to short sleep is linked to higher insulin levels, consistent with the findings from the pooled cross-sectional analysis.

In the ancestry-specific longitudinal analyses, the significant associations observed in the pooled analysis were largely replicated (eTable 4). The positive association between PRS-SS and BMI was replicated in EUR, while the negative association between PRS-SD and BMI was replicated in both EUR and AMR. Several ancestry groups showed significant correlations between PRS-SS, PRS-SD, and glycemic measures. In EUR, a higher genetic risk for short sleep duration was longitudinally linked to increased insulin levels, consistent with the pooled findings. In AFR, short sleep duration was significantly negatively associated with HbA1C, suggesting a lower diabetes risk— a result differing from pooled and other ancestry group findings but consistent with the AFR cross-sectional analysis results. For EAS, longitudinal analysis showed a positive association between longer sleep duration and fasting glucose. No other significant associations between sleep traits PRS and glycemic measures were found in the ancestry-specific longitudinal analyses.

*PRS vs. sleep duration contribution analysis*

The correlation test showed that measured sleep duration was weakly correlated with PRS-SD (β=1.259, $R^2$=0.007) and PRS-SS (β=-29.49, $R^2$=0.002) (eFigure 3). Including measured sleep duration as a covariate in the PRS association analyses rendered all observed associations non-significant in both pooled (Table 2 and Table 3) and ancestry-specific analyses (eTable 5 and eTable 6). Sleep duration demonstrated a relatively stronger contribution compared to PRS in cross-sectional analyses, explaining between 85.6% and 99.9% of the variation when considering only these two factors. In longitudinal analyses, its relative contribution was lower, ranging from 7.1% to 44.0%, where individual variability in outcomes was accounted for (eTable 7). These estimates reflect a direct comparison between PRS and measured sleep duration, without incorporating other potential contributors to variability in health outcomes.

## Discussion

In this study, we demonstrated that individual SNPs previously associated with sleep traits (chronotype, sleep duration and short sleep) were linked to health phenotypes, including obesity, diabetes, cardiovascular, and neurological markers. PRS, serving as a proxy for genetic predisposition to these sleep traits, consistently showed associations with key metabolic health metrics in both cross-sectional and longitudinal analyses. However, when actual sleep duration derived from Fitbit data, which reflects both genetic and non-genetic influences, was included in the analyses, the associations between PRS and health outcomes became non-significant, suggesting that sleep behavior can modulate the influence of genetic sleep predispositions on health outcomes. Non-significant mediation analysis indicated that sleep duration acts as a confounder rather than a mediator, with its relative contribution to health outcomes compared to PRS being stronger in cross-sectional analyses (85.6%-99.9%) than in longitudinal analyses (7.1% to 44.0%, eTable 7). These differences were further supported by additional findings that sleep duration was significantly associated with outcomes, even when PRS was included as a covariate, in cross-sectional analyses but not in longitudinal analyses (eTable 5 and eTable 6).

The chronotype SNP rs1421085 (chr16:53767042), which showed the strongest associations across phenotypes, was negatively associated with most anthropometric measures (BMI, weight, WC, and HC) but positively associated with WHRadjBMI in both meta- and ancestry-specific analyses. This suggests that while the SNP may contribute to lower BMI, it may also influence fat distribution, resulting in a higher WHRadjBMI. Further investigation into the genes associated with sleep-trait SNPs and their involvement in established disease pathway, including FTO, LMOD1[21-24], and MES1[25,26] is discussed in the Supplementary Material.

When examining the aggregate effect of chronotype-associated SNPs using PRS, the associations between chronotype and anthropometric measures observed in the individual SNP analysis were not replicated, likely due to the combined effects of other genetic variants with opposing or neutral impacts. While PRS-C was associated with WC in cross-sectional analysis (Table 2), no associations were found with BMI or markers of diabetes risk. These findings align with those of Jones et al., who reported no genetic correlation between chronotype and BMI, type 2 diabetes, or insulin levels[8]. Considering that many observational studies reported strong associations between chronotype and metabolic dysfunction[27,28], circadian misalignment, rather than genetically correlated chronotype, may play a stronger role in metabolic dysfunction, further supporting the prominent role of non-genetic factors in shaping health outcomes related to sleep traits.

For the sleep trait association with diabetes, the relationship between sleep duration and diabetes risk is well established, with evidence indicating that short sleep negatively affects diabetes risk.[29] For instance, a meta-analysis of prospective observational studies revealed a U-shaped relationship between sleep duration and type 2 diabetes risk, where both short and long sleep durations were associated with increased risk, with the lowest risk observed at 7-8 hours per day.[30] Our analyses on individual SNPs and PRS for short sleep and sleep duration also showed significant associations with markers of diabetes risk. In both cross-sectional and longitudinal analyses, genetic risk for short sleep duration was consistently found to be positively associated with insulin levels, while associations with HbA1C and fasting glucose varied in direction across ancestry groups. However, when actual sleep duration measurements were included in the analysis, most of these associations became non-significant in both pooled and ancestry-specific analyses. This highlights the complex interactions between sleep traits and metabolic outcomes, which involve both genetic and non-genetic factors, such as lifestyle, environmental exposures, and social determinants of health − all of which may vary across ancestry groups.

Beyond the interplay of genetic and non-genetic factors, these variations may also reflect potential biases when generalizing GWAS findings, which were predominantly derived from EUR ancestry populations.[31,32] This underscores the need for more diverse cohorts in GWAS studies to improve the accuracy and equity of genetic research. Such biases likely explain why our analysis yielded more significant results in the EUR group compared to other ancestry groups. This disparity is also attributed to sample size differences, as EUR participants comprised around 50% of the sample (56.2% cross-sectional, 43.7% longitudinal; eTable 3 and 4).

Moreover, data availability varied considerably across phenotype measures. Lab values such as HbA1C, fasting glucose, and insulin were less available compared to BMI and WC measurements, further limiting the scope of ancestry-specific analyses. This discrepancy likely contributed to fewer significant correlations between PRS for sleep traits and glycemic measures compared to anthropometric measures.

However, as the AoU dataset continues to grow toward 1 million diverse participants, our study exploring the interaction between genetic profiles, potential covariates, and ancestry-specific effects holds the potential to advance our understanding of factors influencing overall health and their contribution to health disparities, with a more reliable sample size for both minority groups and phenotype measures.

## Conclusion

This study contributes to the growing body of sleep research by providing a comprehensive analysis of the relationship between genetic susceptibility to sleep traits and a wide range of health conditions in the U.S. population. Our findings emphasize the importance of achieving adequate sleep duration, as the adverse associations between shorter sleep and health outcomes were no longer observed when actual sleep duration behavior was included in the models. These findings highlight the potential for adequate sleep to mitigate the health risks associated with a genetic predisposition to shorter sleep duration.

**Acknowledgements**

This work was supported by National Cancer Institute of the National Institutes of Health under K25CA267052 (J.P.). Dr. St-Onge is funded by R35HL155670, R01HL142648, and R01DK128154; Drs. St-Onge and Barragán received funding from AHA grant 24IVPHAA1293965 to support this work. Dr. Barragán received additional funding from CIBER training program support (Instituto de Salud Carlos III). We gratefully acknowledge All of Us participants for their contributions, without whom this research would not have been possible. We also thank the National Institutes of Health's All of Us Research Program for making available the participant data examined in this study.

**Competing interest**

None

**Data availability**

To ensure privacy of participants, data used for this study are available to approved researchers following registration, completion of ethics training and attestation of a data use agreement through the All of Us Research Workbench platform, which can be accessed via https://workbench.researchallofus.org/.

## Code availability

Code used for this study can be made immediately available to any approved researchers on the All of Us Research Workbench platform by contacting our study team.

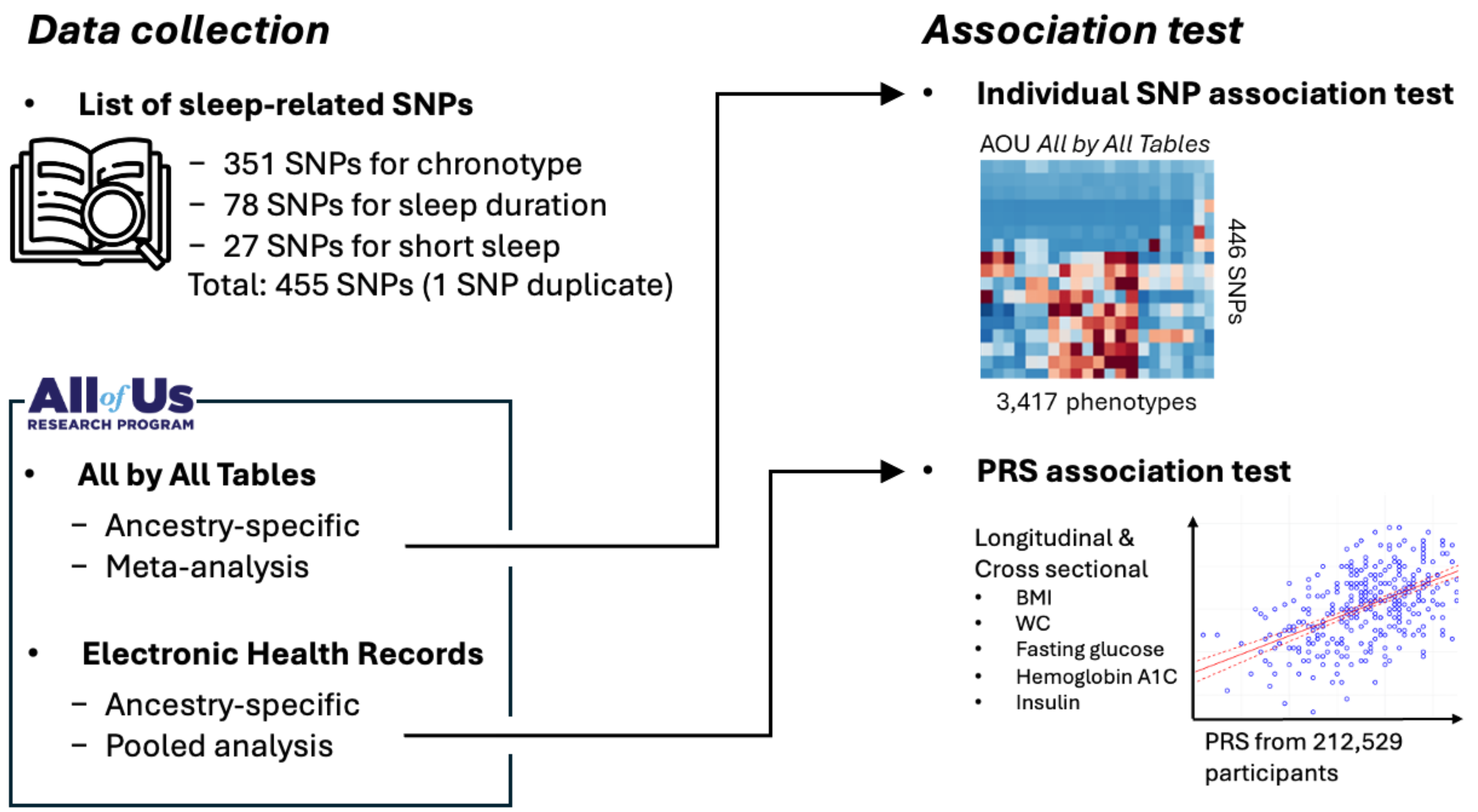

**Figure 1. Study analysis workflow.** We collected a list of sleep-related SNPs from the literature and performed individual SNP association tests using the All by All tables from the AOU database, which provides ancestry-specific results and a meta-analysis combining these results. We then calculated PRS for these SNPs and conducted association tests with both cross-sectional and longitudinal measurements of BMI, waist circumference (WC), fasting glucose, hemoglobin A1C, and insulin.

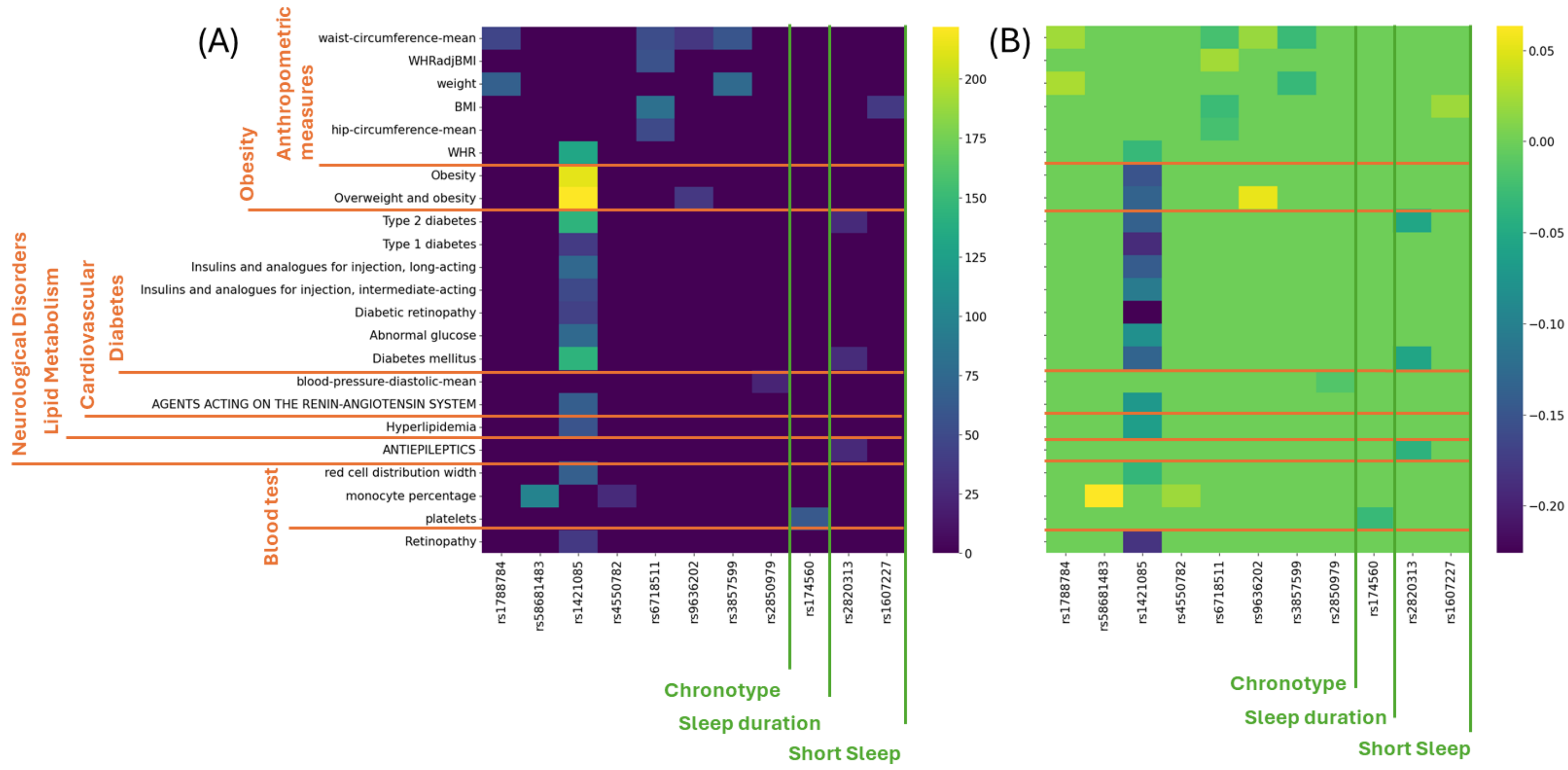

**Figure 2. GWAS meta-analysis**. We identified 23 unique phenotypes associated with 11 sleep-trait-associated SNPs. The 23 phenotypes are categorized into eight health conditions: anthropometric measures, obesity, diabetes, cardiovascular, lipid metabolism, sleep disorders, and blood tests (eTable 2). Additionally, chronotype showed an association with the phenotype 'retinopathy' outside of the eight categories. (A) The colormap represents the median -log(p-value) from GWAS across the entire population, including analyses with all ancestries and leave-one-out analyses. (B) The colormap displays the corresponding effect sizes (the median of beta values) for each association.

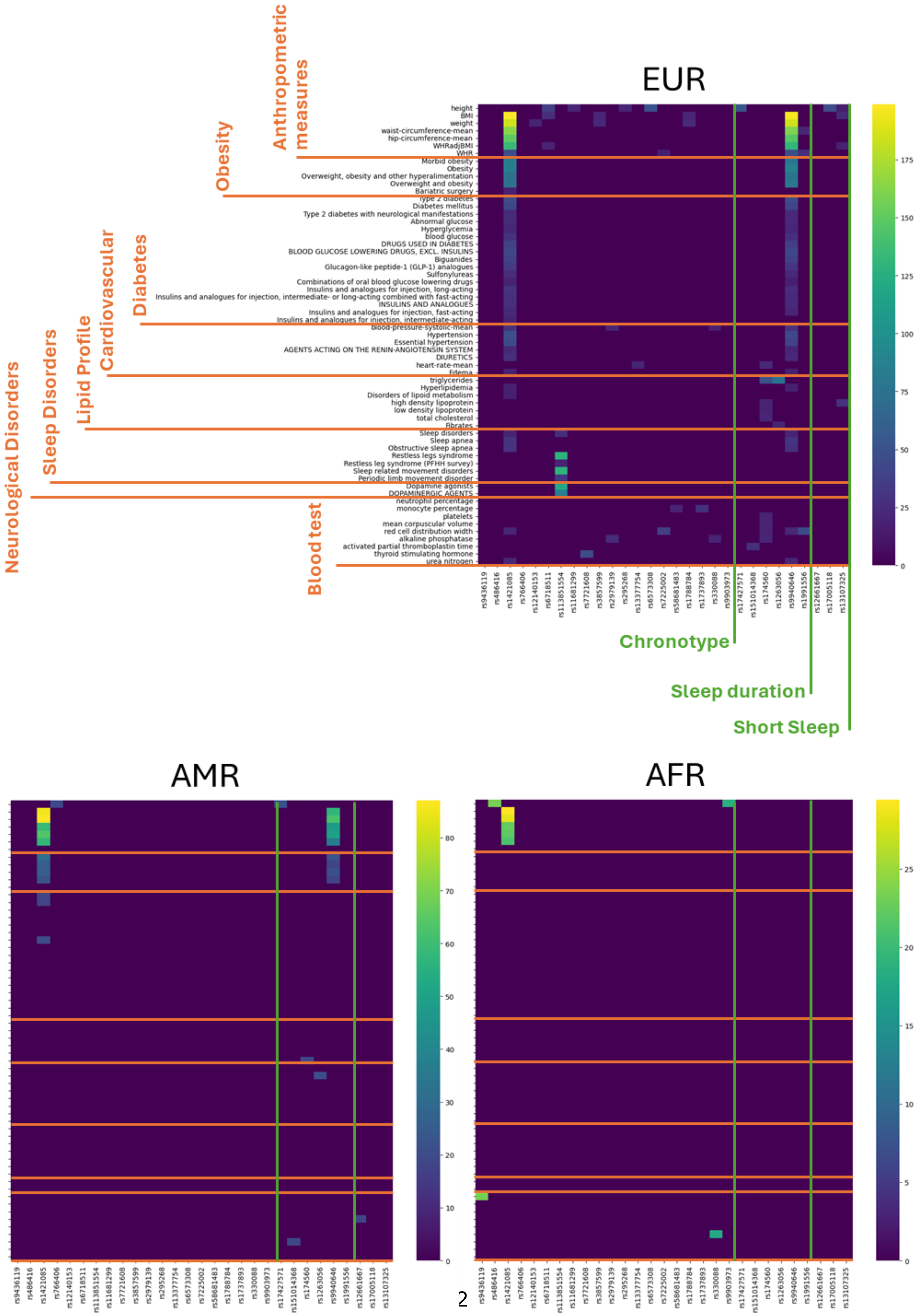

EUR

Anthropometric measures
Obesity
Diabetes
Cardiovascular
Lipid Profile
Sleep Disorders
Neurological Disorders
Blood test

height
BMI
weight
waist-circumference-mean
hip-circumference-mean
WHRadjBMI
WHR
Morbid obesity
Obesity
Overweight, obesity and other hyperalimentation
Overweight and obesity
Bariatric surgery
Type 2 diabetes
Diabetes mellitus
Type 2 diabetes with neurological manifestations
Abnormal glucose
Hyperglycemia
blood glucose
DRUGS USED IN DIABETES
BLOOD GLUCOSE LOWERING DRUGS, EXCL. INSULINS
Biguanides
Glucagon-like peptide-1 (GLP-1) analogues
Sulfonylureas
Combinations of oral blood glucose lowering drugs
Insulins and analogues for injection, long-acting
Insulins and analogues for injection, intermediate- or long-acting combined with fast-acting
INSULINS AND ANALOGUES
Insulins and analogues for injection, fast-acting
Insulins and analogues for injection, intermediate-acting
blood-pressure-systolic-mean
Hypertension
Essential hypertension
AGENTS ACTING ON THE RENIN-ANGIOTENSIN SYSTEM
DIURETICS
heart-rate-mean
Edema
triglycerides
Hyperlipidemia
Disorders of lipoid metabolism
high density lipoprotein
low density lipoprotein
total cholesterol
Fibrates
Sleep disorders
Sleep apnea
Obstructive sleep apnea
Restless legs syndrome
Restless leg syndrome (PFHH survey)
Sleep related movement disorders
Periodic limb movement disorder
Dopamine agonists
DOPAMINERGIC AGENTS
neutrophil percentage
monocyte percentage
platelets
mean corpuscular volume
red cell distribution width
alkaline phosphatase
activated partial thromboplastin time
thyroid stimulating hormone
urea nitrogen

rs9436119
rs486416
rs1421085
rs766406
rs12140153
rs6718511
rs113851554
rs11681299
rs7721608
rs3857599
rs2979139
rs295268
rs13377754
rs6573308
rs7225002
rs58681483
rs1788784
rs1737893
rs330088
rs9903973
rs17427571
rs151014368
rs174560
rs1263056
rs9940646
rs1991556
rs12661667
rs17005118
rs13107325

0
25
50
75
100
125
150
175

Chronotype
Sleep duration
Short Sleep

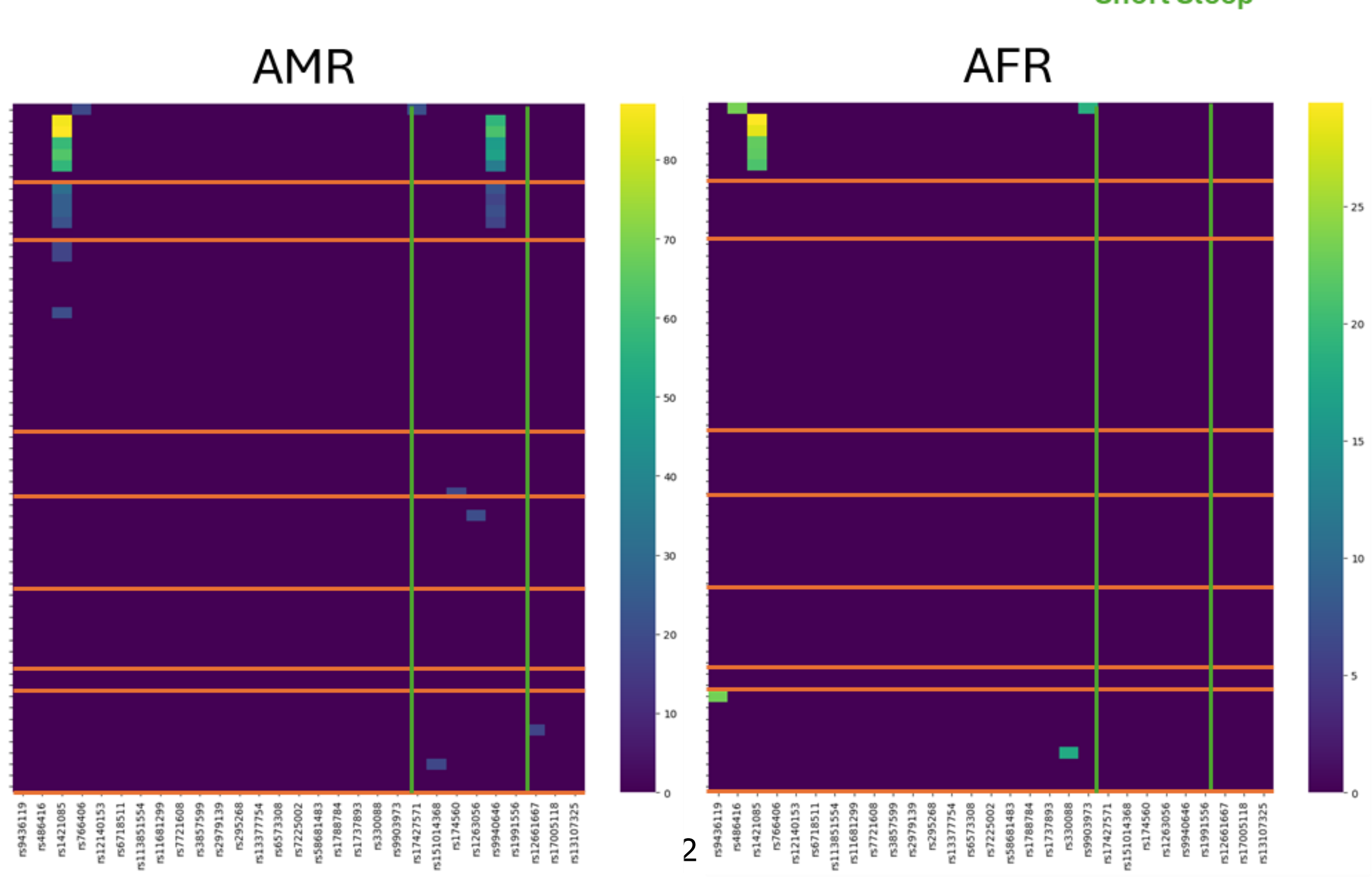

**Figure 3. GWAS by ancestry groups**. We identified 61 unique phenotypes associated with 29 sleep-trait-associated SNPs within the EUR, AFR, and AMR ancestry groups. The 61 phenotypes are categorized into eight health conditions: anthropometric measures, obesity, diabetes, cardiovascular, lipid metabolism, sleep disorders, and blood tests (eTable 2). The colormap represents -log(p-value). A heatmap showing the effect sizes for each association is provided in eFigure 2.

**Table 1. Characteristics of AoU participants included in the analysis.**

| | | N | % |
|---|---|---|---|
| Number of participants with genomic data | | 245,394 | |
| Number of participants with genomic & PM or EHR data meeting the inclusion criteria | | 212,529 | |
| Number of participants with available Fitbit data within the 219,061 | | 7,655 | |
| Sex | Female | 126,144 | 59.4 |
| | Male | 86,385 | 40.6 |
| Age, median (IQR) | | 58 (42-70) | |
| Genetic ancestry | European (EUR) | 122,398 | 55.9 |
| | African (AFR) | 48,193 | 22.0 |
| | Admixed American (AMR) | 39,375 | 18.0 |
| | East Asian (EAS) | 5,387 | 2.5 |
| | South Asian (SAS) | 2,857 | 1.3 |
| | Middle Eastern (MID) | 851 | 0.4 |
| Ethnicity | Not Hispanic | 176,239 | 80.5 |
| | Hispanic | 42,822 | 19.5 |

PM: physical measurements; EHR: electronic health records

**Table 2. Pooled cross-sectional analysis results.**

| | PRS-C | | | PRS-C with Fitbit† | | | PRS-SS | | | PRS-SS with Fitbit† | | | PRS-SD | | | PRS-SD with Fitbit† | | |
|---|---|---|---|---|---|---|---|---|---|---|---|---|---|---|---|---|---|---|
| | β | 95% CI | p-value | β | 95% CI | p-value | β | 95% CI | p-value | β | 95% CI | p-value | β | 95% CI | p-value | β | 95% CI | p-value |
| BMI | 0.002 | [-0.003, 0.007] | 0.424 | -0.006 | [-0.029, 0.017] | 0.599 | 0.012*** | [0.008, 0.017] | <0.001 | 0.012 | [-0.01, 0.034] | 0.281 | -0.022*** | [-0.027, -0.017] | <0.001 | -0.018 | [-0.041, 0.005] | 0.127 |
| WC | 0.005* | [0.001, 0.01] | 0.014 | 0.003 | [-0.017, 0.023] | 0.760 | 0.011*** | [0.007, 0.015] | <0.001 | 0.005 | [-0.014, 0.025] | 0.587 | -0.016*** | [-0.02, -0.011] | <0.001 | -0.001 | [-0.021, 0.019] | 0.936 |
| Fasting glucose | -0.003 | [-0.026, 0.02] | 0.789 | -0.041 | [-0.141, 0.058] | 0.415 | -0.007 | [-0.028, 0.014] | 0.529 | -0.037 | [-0.131, 0.057] | 0.437 | 0.001 | [-0.022, 0.024] | 0.919 | 0.031 | [-0.066, 0.128] | 0.531 |
| HbA1C | -0.001 | [-0.008, 0.006] | 0.794 | 0.010 | [-0.025, 0.046] | 0.568 | 0.004 | [-0.002, 0.011] | 0.162 | -0.027 | [-0.061, 0.007] | 0.119 | -0.007 | [-0.014, 0.0] | 0.055 | 0.018 | [-0.018, 0.054] | 0.316 |
| Insulin | 0.005 | [-0.054, 0.064] | 0.865 | 0.085 | [-0.241, 0.41] | 0.603 | 0.050 | [-0.004, 0.104] | 0.070 | 0.089 | [-0.201, 0.379] | 0.541 | -0.070* | [-0.133, -0.006] | 0.031 | 0.137 | [-0.161, 0.436] | 0.360 |

Coefficients (β) from ordinary least squares (OLS) multivariable linear regression analysis of PRS for chronotype, short sleep, sleep duration on BMI, WC, fasting glucose, insulin, and HbA1C, adjusted for age, ancestry, ethnicity, sex, smoking, obesity, and diabetes. PRS and all measurements were standardized to a mean of 0 and a standard deviation of 1. Significance levels: *p < 0.05, **p < 0.01, ***p < 0.001. †Association test with Fitbit-measured sleep duration included as a covariate.

**Table 3. Pooled longitudinal analysis results.**

| | PRS-C | | | PRS-C with Fitbit† | | | PRS-SS | | | PRS-SS with Fitbit† | | | PRS-SD | | | PRS-SD with Fitbit† | | |
|---|---|---|---|---|---|---|---|---|---|---|---|---|---|---|---|---|---|---|
| | β | 95% CI | p-value | β | 95% CI | p-value | β | 95% CI | p-value | β | 95% CI | p-value | β | 95% CI | p-value | β | 95% CI | p-value |
| BMI | 0.002 | [-0.007, 0.011] | 0.637 | 0.006 | [-0.036, 0.047] | 0.790 | 0.017*** | [0.008, 0.025] | <0.001 | 0.025 | [-0.015, 0.065] | 0.225 | -0.016*** | [-0.026, -0.007] | <0.001 | -0.041 | [-0.082, 0.001] | 0.053 |
| WC | N/A | N/A | N/A | N/A | N/A | N/A | N/A | N/A | N/A | N/A | N/A | N/A | N/A | N/A | N/A | N/A | N/A | N/A |
| Fasting glucose | 0.009 | [-0.023, 0.041] | 0.573 | -0.138 | [-0.329, 0.052] | 0.155 | -0.016 | [-0.044, 0.012] | 0.262 | 0.008 | [-0.136, 0.152] | 0.913 | 0.028 | [-0.004, 0.061] | 0.084 | 0.053 | [-0.134, 0.24] | 0.578 |
| HbA1C | -0.003 | [-0.011, 0.005] | 0.469 | -0.006 | [-0.044, 0.032] | 0.754 | 0.003 | [-0.005, 0.011] | 0.430 | -0.009 | [-0.045, 0.027] | 0.625 | -0.005 | [-0.013, 0.004] | 0.290 | -0.004 | [-0.042, 0.035] | 0.851 |
| Insulin | 0.091 | [-0.042, 0.224] | 0.181 | 0.287 | [-0.375, 0.948] | 0.396 | 0.134* | [0.008, 0.259] | 0.037 | 0.140 | N/A | N/A | -0.007 | [-0.109, 0.095] | 0.890 | -0.082 | [-0.411, 0.247] | 0.626 |

Fixed effects coefficients (β) from longitudinal analyses of PRS for sleep traits on BMI, WC, fasting glucose, HbA1C, insulin. PRS and all measurements were standardized to a mean of 0 and a standard deviation of 1. The linear mixed models were adjusted for ancestry, ethnicity, sex, smoking, obesity, and diabetes. Significance levels: *$p < 0.05$, **$p < 0.01$, ***$p < 0.001$. †Association test with Fitbit-measured sleep duration included as a covariate.

Supplementary Materials

**AoU recruitment and enrollment.** The All of Us (AoU) Research program provides a large, comprehensive, and diverse biomedical dataset of participants in the United States. The goal of this program is to accelerate medical research and breakthroughs by enrolling at least one million individuals[10]. AoU began in May 2018 and, as of August 2024, more than 826,000 participants have been enrolled with over 80% from historically underrepresented groups in biomedical research. Participants provide AoU with a diverse set of data from Electronic Health Records (EHR), surveys, physical measurements, biospecimens, and wearable devices[15]. These data are organized into Curated Data Repositories (CDRs), which are divided into different access tiers based on the level of data detail available to researchers in the AoU Researcher Workbench. For our analyses, we used the Controlled Tier Dataset v7 CDR, which includes data from 413,457 participants enrolled from May 2018 to July 2022[10,14,15]. Data from participants enrolled after July 2022 were not available as of August 2024.

**AoU datasets.** We utilized five datasets from AoU Research Program – Genomic data, one-time Physical Measurement data, longitudinal EHR data, Fitbit sleep data, and All by All tables. *Genomic data:* The AoU's Controlled Tier provides access to participant genomic data in various formats. Our analyses focused on the short-read whole genome sequencing (srWGS) SNP and Indel callset, stored in the Hail VariantDataset (VDS) format. Using the provided table of samples flagged for relatedness, we removed 15,375 related participants from the 245,394 total participants with genomic data. Of the remaining 230,019 participants, demographic data were available for 230,013 participants. We excluded data from 4,713 participants (2.0%) with sex at birth other than Male or Female and 8,806 participants (3.8%) with ethnicity other than Hispanic or non-Hispanic, leaving 212,529 participants (Table 1, eFigure 1).

*Physical Measurement and EHR data:* The Physical Measurement dataset contains one-time measurements taken at enrollment, while longitudinal data are available through the EHR dataset. Both datasets are standardized to the Observational Medical Outcomes Partnership (OMOP) common data model (CDM), in which medical information such as diagnoses, drugs, and procedures are mapped to standard concepts. We used AoU's Cohort Builder tool in the Researcher Workbench to filter participants with available WGS and EHR data. With this cohort, we pulled longitudinal data for hemoglobin A1C (HbA1C), fasting glucose, and insulin lab results, as well as height, weight, and waist circumference (WC) measurements from the EHR data using relevant concept IDs (eFigure 1). Additionally, one-time height, weight, and WC measurements taken at enrollment were extracted from the Physical Measurement (PM) dataset and pooled with the EHR data. BMI values were calculated from the height and weight measurements for each participant.

*Fitbit sleep data:* We used Fitbit-derived sleep duration data to calculate average sleep duration for participants with available Fitbit data in AoU. Of the 212,529 participants in our cohort, 7,655 had available sleep duration data. Daily sleep duration was calculated by summing up the minutes asleep for any sleep period classified as the 'main sleep' of the day. These values were then averaged to derive the average daily sleep duration for each participant.

*Exclusion criteria for measurements:* All lab and anthropometric measurements were filtered to exclude biologically implausible values. The following ranges were considered biologically implausible: >50% for HbA1C, >1000 mg/dL for fasting glucose, >100 µIU/mL for insulin, <111.8 cm or >228.6 cm for height[11,12], <24.9 kg or >453.6 kg for weight[11,12], and <5 kg/m$^2$ or

≥100 kg/m$^2$ for BMI[11,13]. To improve reliability of the BMI data, height and weight measurements underwent quality control following established methods in literature. In addition to filtering out implausible measurements, these methods included removing any height and weight measurements taken during inpatient and emergency department visits and any measurements taken within one year of a pregnancy-related diagnosis.[11,13] We then removed within-subject outliers, which were measurements for each participant that met the following criteria outlined by Cheng et al.[12]: weight measurements where (1) the range was >22.7 kg and the absolute difference between that specific weight and average weight was >70% of the range or (2) the standard deviation (SD) was >20% of the average weight and the absolute difference between that particular weight and average weight was greater than the SD; and height measurements where (1) the absolute difference between that particular height and average height was greater than the SD and (2) the SD was >2.5% of the average height. Any participants that still had a >5cm height change were removed. The remaining height and weight values were matched by measurement date and used to calculate BMI values for each participant. Finally, for all lab and anthropometric measurements, the median was used if a participant had multiple measurements on the same day. The full filtering process including participant counts can be seen in eFigure 1.

*All by All Tables:* As of June 2024, All by All tables were released to Controlled Tier users in the AoU Researcher Workbench. Leveraging srWGS and phenotypic data from ~250,000 participants in the AoU Researcher Workbench, these tables contain GWAS results for 8,895 high-quality ancestry group and phenotype pairs. The ancestry groups included were African (AFR), Admixed American (AMR), European (EUR), East Asian (EAS), South Asian (SAS), and Middle Eastern (MID). In total, there were 3,414 unique phenotypes of six categories:

physical measurements, lab measurements, phecodes, phecodeX, personal and family health history, and EHR sourced drugs and medications. In addition to results by ancestry, results from a meta-analysis that combined ancestry-specific results were also available.

**PRS calculation.** Weights used in the PRS calculation were the reported effect sizes from published GWAS, which were odds ratios (OR) for morning chronotype[8], OR for short sleep[9], and beta values representing 1-minute increments for sleep duration.[9] OR values were transformed to log OR prior to calculation. Because the reported SNP locations were relative to the GRCh37 reference genome and AoU dataset uses GRCh38, we employed the pyliftover Python package to convert the GRCh37 positions to GRCh38. Utilizing the AoUPRS package[19], GWAS SNPs were matched to those in AoU by genomic location and PRS was calculated for each sleep trait by summing the number of effect alleles in an individual weighted by their effect size.

**Formulas used in cross-sectional and longitudinal analyses**

*Cross-sectional analysis:* We performed multivariable linear regression for each pair of sleep trait PRS and measurement as follows, where β represents the coefficients and ε denotes the error term:

$$\text{Measurement} = \beta_0 + \beta_1(\text{PRS}) + \beta_2(\text{smoking}) + \beta_3(\text{obesity}) + \beta_4(\text{diabetes}) + \beta_5(\text{age}) + \beta_6(\text{ancestry}) + \beta_7(\text{ethnicity}) + \beta_8(\text{sex}) + \varepsilon$$

All continuous variables were standardized to a mean of 0 and a standard deviation of 1. We assessed multicollinearity among independent variables using the variance inflation factor (VIF) and iteratively excluded covariates with high VIF until all variables had VIF less than 10.

*Longitudinal analysis:* We employed linear mixed models to account for the correlation between longitudinal measurements within individuals while assessing the association between each sleep trait PRS and measurement pair. In this linear mixed models, we included random intercepts to account for individual-level variation in baseline outcomes and random slopes for age at measurement, allowing the relationship between age at measurement and each outcome to vary across participants. PRS was modeled as a fixed effect and, along with covariates, was estimated via conditional likelihood. The formulation is as follows, where $\mathrm{b}_{0i}$ is the random intercept and $\mathrm{b}_{1i}$ is the random slope for each participant *i*.

$$\begin{aligned}\mathrm{Measurement}_{\mathrm{ij}} &= \beta_0 + \beta_1(\text{age at measurement}_{ij}) + \beta_2(\mathrm{PRS}_i) + \beta_3(\mathrm{smoking}_i) \\ &+ \beta_4(\mathrm{obesity}_i) + \beta_5(\mathrm{diabetes}_i) + \beta_6(\mathrm{ancestry}_i) + \beta_7(\mathrm{ethnicity}_i) + \beta_8(\mathrm{sex}_i) \\ &+ b_{0i} + b_{1i}(\text{age at measurement}_{ij}) + \varepsilon_{ij}\end{aligned}$$

We included the same covariates in the longitudinal analyses as in the cross-sectional analyses. Standardization of continuous variables and handling of multicollinearity aligned with the cross-sectional analyses as well.

**Genetic associations with sleep traits and their implications for disease pathways**

We examined the genes associated with specific SNPs linked to various health phenotypes in our study and investigated their roles in established disease pathways. For example, the chronotype SNP rs1421085 (chr16:53767042) is located within the FTO gene region, a well-known predictor of polygenic obesity—a significant risk factor for type 2 diabetes and other metabolic disorders. Similarly, the sleep duration SNP rs9940646 (chr16:53766717), which in our analysis was associated with phenotypes related to obesity, diabetes, and cardiovascular conditions, is also situated within the FTO gene region, further highlighting FTO's central role in obesity-related traits. Another notable example is the short sleep SNP rs2820313 (chr1:201901093), a mutation in the leiomodin1 (LMOD1) gene, a smooth muscle-restricted and relatively understudied gene.[19] In our study, this SNP demonstrated significant associations with the use of antiepileptic medications, BMI, and type 2 diabetes (Figure 2). These findings align with the known relationship between sleep patterns and seizure activity, as sleep deprivation can trigger seizures in individuals with epilepsy[20] – a condition often associated with obesity and diabetes.[21] Additionally, recent studies have implicated LMOD1 in nodding syndrome, a form of pediatric epilepsy, suggesting that antibodies against LMOD1 may contribute to its pathogenesis.[22] The chronotype SNP rs113851554 (chr2:66523432), located in the MES1 gene, showed a significant association with sleep disorders and dopamine-related medications (e.g., dopamine agonists, dopaminergic agents), but only within the EUR ancestry group. Previous studies have reported that the MES1 GWAS signals were among the strongest genetic associations reported for the development of restless legs syndrome, a common sleep-related disorder.[23,24]

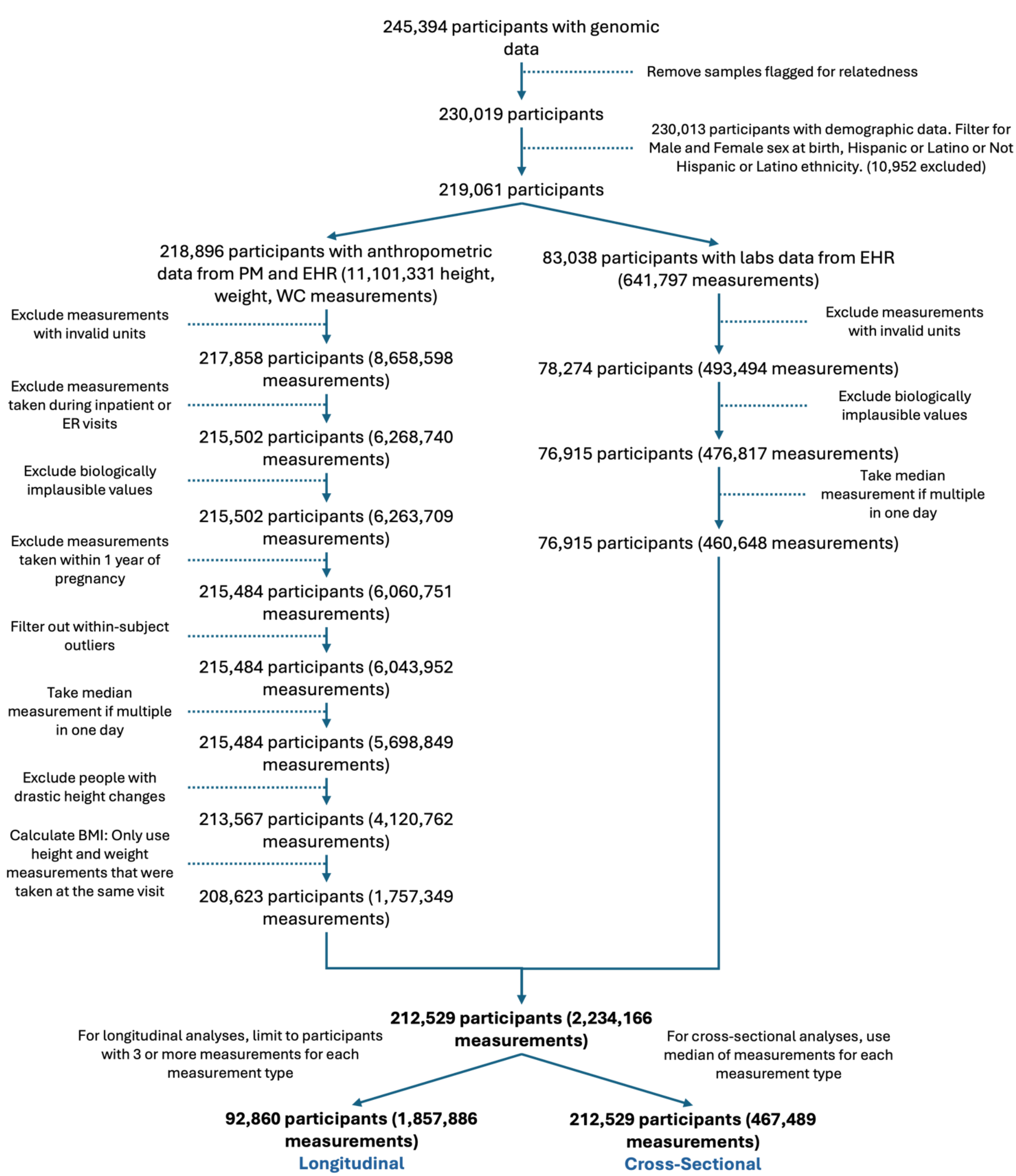

**eFigure 1. Participant inclusion criteria flow diagram.** A flow diagram showing the filtering criteria and the number of participants and measurements remaining after each step.

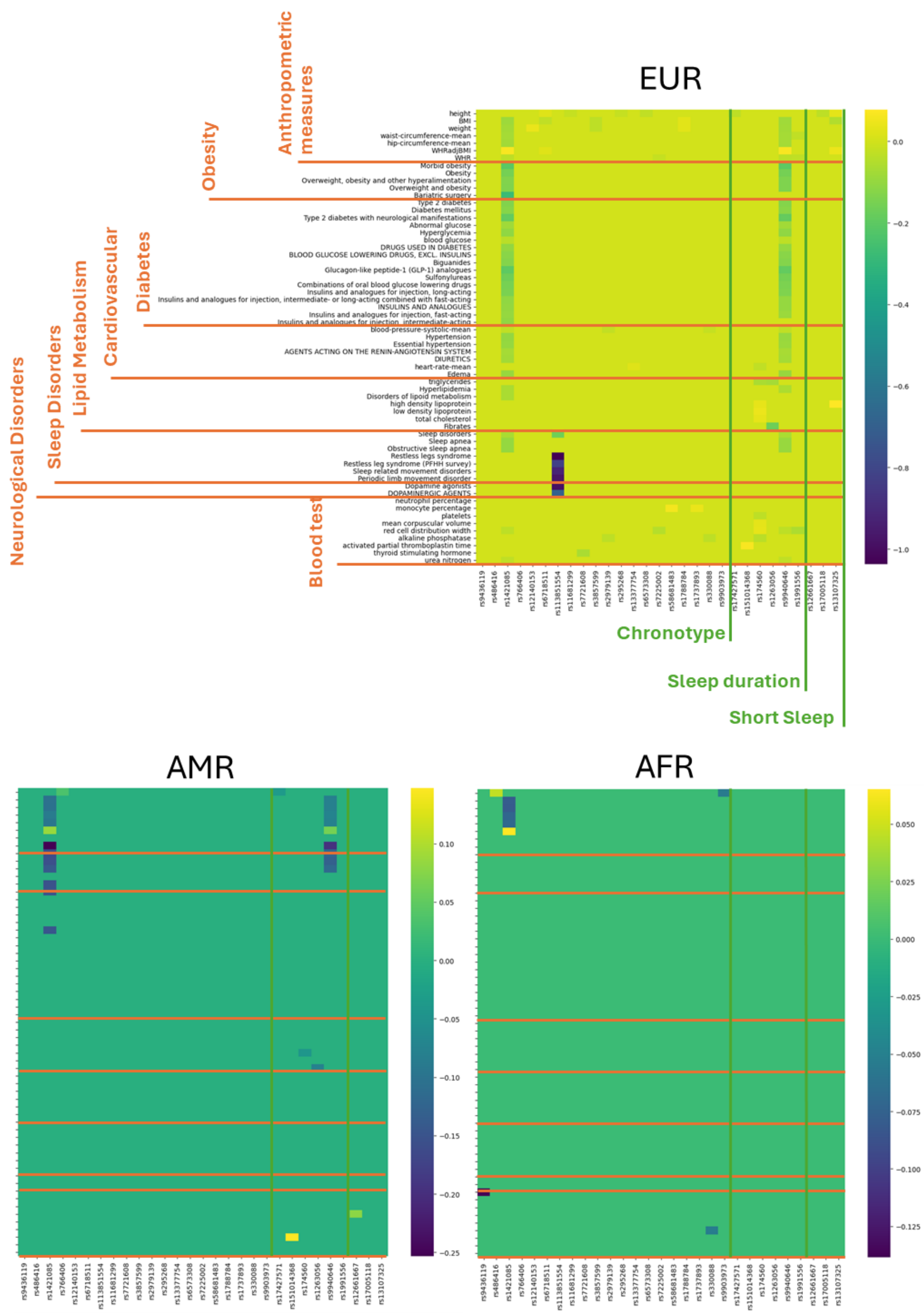

**eFigure 2. GWAS by ancestry groups.** A heatmap presenting the effect sizes for associations between sleep trait SNPs and phenotypes.

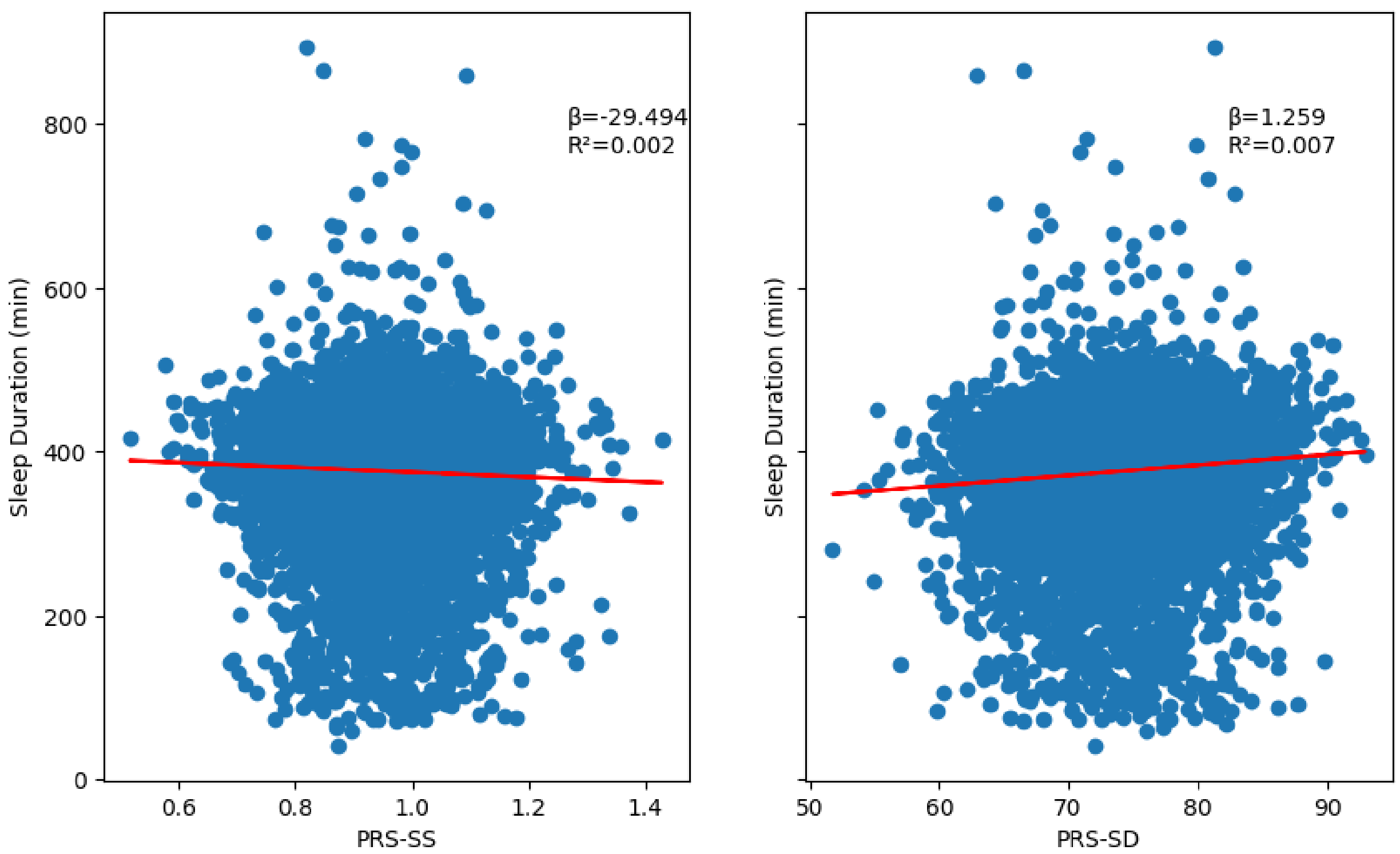

**eFigure 3.** Correlation between PRS-SS and actual sleep duration measurements from Fitbit data (left) and between PRS-SD and actual sleep duration measurements from Fitbit data (right).

## eTable 1. List of concept IDs, concept names, and ICD codes used to query data.

| Measurement/Condition | Query Value |
|---|---|
| Height | Concept ID: 903133, 3019171, 3023540, 3036277 |
| Weight | Concept ID: 903121, 3013762, 3023166, 3025315, 3027492 |
| Waist Circumference | Concept ID: 903124 |
| Hemoglobin A1C | Concept ID: 300309, 3004410, 3005673, 3007263, 4184637, 4197971, 40762352, 42869630 |
| Fasting glucose | Concept ID: 3036671, 3037110, 3037187, 4182052, 46235168 |
| Insulin | Concept ID: 3016244, 3022466, 36303996 |
| Obesity | Concept ID: 433736 and all descendant concepts |
| Diabetes | Concept ID: 201820 and all descendant concepts<br><br>Concept names for antidiabetic drugs:<br>• Type 1 diabetes: insulin, pramlintide, insulin glulisine, insulin lispro, insulin aspart, insulin glargine, insulin detemir, insulin degludec, insulin NPH<br>• Type 2 diabetes: acetohexamide, tolazamide, chlorpropamide, glipizide, glyburide, glimepiride, repaglinide, nateglinide, metformin, rosiglitazone, pioglitazone, troglitazone, acarbose, miglitol, sitagliptin, exenatide, saxagliptin, linagliptin, liraglutide, semaglutide, canagliflozin, dapagliflozin, empagliflozin, alogliptin, colesevelam, albiglutide, dulaglutide, lixisenatide |
| Smoking | Concept ID: 440612, 4146763, 4103418, 4209423, 37110444, 37109023, 437264, 36716478, 765451, 3654548, 36716473 and all descendant concepts |
| Pregnancy | ICD-9 Code: 630-679.14<br><br>ICD-10 Code: O0.0-O9A53, Z33.X, Z34.X, Z3A.X, Z37.X, Z38.X, Z39.0, A34.X |

## eTable 2. Grouping of phenotypes associated with sleep trait SNPs identified in meta-analysis and ancestry-specific analyses

| Phenotype group | Phenotypes | |
|---|---|---|
| | **Meta-analysis** | **Ancestry-specific analysis** |
| Anthropometric Measures | 'waist-circumference-mean', 'WHRadjBMI', 'weight','BMI', 'hip-circumference-mean', 'WHR' | 'height', 'BMI', 'weight', 'waist-circumference-mean', 'hip-circumference-mean', 'WHRadjBMI', 'WHR' |
| Obesity and Related Conditions | 'Obesity', 'Overweight and obesity' | 'Morbid obesity', 'Obesity', 'Overweight, obesity and other hyperalimentation', 'Overweight and obesity', 'Bariatric surgery', |
| Diabetes and Blood Glucose Management | 'Type 2 diabetes', 'Type 1 diabetes', 'Insulins and analogues for injection, long-acting', 'Insulins and analogues for injection, intermediate-acting', 'Diabetic retinopathy', 'Abnormal glucose', 'Diabetes mellitus' | 'Type 2 diabetes', 'Diabetes mellitus', 'Type 2 diabetes with neurological manifestations', 'Abnormal glucose', 'Hyperglycemia', 'blood glucose', 'DRUGS USED IN DIABETES', 'BLOOD GLUCOSE LOWERING DRUGS, EXCL. INSULINS', 'Biguanides', 'Glucagon-like peptide-1 (GLP-1) analogues', 'Sulfonylureas', 'Combinations of oral blood glucose lowering drugs', 'Insulins and analogues for injection, long-acting', 'Insulins and analogues for injection, intermediate- or long-acting combined with fast-acting', 'INSULINS AND ANALOGUES', 'Insulins and analogues for injection, fast-acting', 'Insulins and analogues for injection, intermediate-acting' |
| Cardiovascular and Hypertension-Related | 'blood-pressure-diastolic-mean', 'AGENTS ACTING ON THE RENIN-ANGIOTENSIN SYSTEM' | 'blood-pressure-systolic-mean', 'Hypertension', 'Essential hypertension', 'AGENTS ACTING ON THE RENIN-ANGIOTENSIN SYSTEM', 'DIURETICS', 'heart-rate-mean', 'Edema' |
| Lipid Metabolism | 'Hyperlipidemia' | 'triglycerides', 'Hyperlipidemia', 'Disorders of lipoid metabolism', 'high density lipoprotein', 'low density lipoprotein', 'total cholesterol', 'Fibrates' |
| Sleep Disorders | NA | 'Sleep disorders', 'Sleep apnea', 'Obstructive sleep apnea', 'Restless legs syndrome', 'Restless leg syndrome (PFHH survey)', 'Sleep related movement disorders', 'Periodic limb movement disorder' |

| Phenotype group | Phenotypes | |
|---|---|---|
| | **Meta-analysis** | **Ancestry-specific analysis** |
| Neurological Disorders and Medications | 'ANTIEPILEPTICS' | 'Dopamine agonists', 'DOPAMINERGIC AGENTS' |
| Hematological and Laboratory Values | 'red cell distribution width', 'monocyte percentage', 'platelets' | 'neutrophil percentage', 'monocyte percentage', 'platelets', 'mean corpuscular volume', 'red cell distribution width', 'alkaline phosphatase', 'activated partial thromboplastin time', 'thyroid stimulating hormone', 'urea nitrogen' |

## eTable 3. Cross-sectional analyses results by ancestry

| Genetic Ancestry | N | % | Measurement | N | % | PRS-C β | PRS-C 95% CI | PRS-C p-value | PRS-SS β | PRS-SS 95% CI | PRS-SS p-value | PRS-SD β | PRS-SD 95% CI | PRS-SD p-value |
|---|---|---|---|---|---|---|---|---|---|---|---|---|---|---|
| All | 212,529 | 100.0% | BMI | 190,876 | 89.8% | 0.002 | [-0.003, 0.007] | 0.424 | 0.012*** | [0.008, 0.017] | <0.001 | -0.022*** | [-0.027, -0.017] | <0.001 |
| | | | WC | 192,979 | 90.8% | 0.005* | [0.001, 0.01] | 0.014 | 0.011*** | [0.007, 0.015] | <0.001 | -0.016*** | [-0.02, -0.011] | <0.001 |
| | | | Fasting glucose | 7,372 | 3.5% | -0.003 | [-0.026, 0.02] | 0.789 | -0.007 | [-0.028, 0.014] | 0.529 | 0.001 | [-0.022, 0.024] | 0.919 |
| | | | HbA1C | 75,055 | 35.3% | -0.001 | [-0.008, 0.006] | 0.794 | 0.004 | [-0.002, 0.011] | 0.162 | -0.007 | [-0.014, 0.0] | 0.055 |
| | | | Insulin | 1,207 | 0.6% | 0.005 | [-0.054, 0.064] | 0.865 | 0.050 | [-0.004, 0.104] | 0.070 | -0.070* | [-0.133, -0.006] | 0.031 |
| EUR | 119,539 | 56.2% | BMI | 106,895 | 50.3% | 0.003 | [-0.003, 0.009] | 0.339 | 0.016*** | [0.01, 0.022] | <0.001 | -0.016*** | [-0.022, -0.01] | <0.001 |
| | | | WC | 108,451 | 51.0% | 0.007** | [0.002, 0.012] | 0.004 | 0.013*** | [0.008, 0.018] | <0.001 | -0.014*** | [-0.019, -0.009] | <0.001 |
| | | | Fasting glucose | 4,413 | 2.1% | -0.004 | [-0.031, 0.023] | 0.761 | 0.003 | [-0.023, 0.03] | 0.801 | 0.008 | [-0.019, 0.034] | 0.579 |
| | | | HbA1C | 43,747 | 20.6% | 0.006 | [-0.003, 0.014] | 0.176 | 0.008 | [-0.001, 0.016] | 0.067 | -0.007 | [-0.015, 0.002] | 0.114 |
| | | | Insulin | 624 | 0.3% | 0.024 | [-0.051, 0.099] | 0.533 | 0.066 | [-0.009, 0.14] | 0.084 | -0.066 | [-0.14, 0.009] | 0.085 |
| AFR | 47,212 | 22.2% | BMI | 42,737 | 20.1% | -0.007 | [-0.016, 0.002] | 0.108 | 0.002 | [-0.007, 0.011] | 0.659 | -0.020*** | [-0.029, -0.011] | <0.001 |
| | | | WC | 43,828 | 20.6% | 0.000 | [-0.009, 0.008] | 0.916 | 0.005 | [-0.003, 0.014] | 0.245 | -0.009* | [-0.017, -0.0] | 0.047 |
| | | | Fasting glucose | 1,397 | 0.7% | 0.021 | [-0.027, 0.069] | 0.397 | -0.023 | [-0.071, 0.025] | 0.347 | -0.012 | [-0.061, 0.036] | 0.622 |
| | | | HbA1C | 16,140 | 7.6% | -0.005 | [-0.019, 0.008] | 0.451 | -0.017* | [-0.031, -0.004] | 0.013 | 0.006 | [-0.008, 0.02] | 0.387 |
| | | | Insulin | 371 | 0.2% | -0.008 | [-0.109, 0.094] | 0.881 | 0.036 | [-0.065, 0.137] | 0.487 | -0.010 | [-0.113, 0.093] | 0.851 |
| AMR | 36,965 | 17.4% | BMI | 33,037 | 15.5% | 0.012* | [0.001, 0.023] | 0.026 | 0.009 | [-0.001, 0.02] | 0.080 | -0.032*** | [-0.043, -0.021] | <0.001 |
| | | | WC | 32,721 | 15.4% | 0.005 | [-0.004, 0.015] | 0.284 | 0.008 | [-0.002, 0.018] | 0.116 | -0.021*** | [-0.03, -0.011] | <0.001 |
| | | | Fasting glucose | 1,299 | 0.6% | -0.022 | [-0.072, 0.028] | 0.393 | -0.026 | [-0.076, 0.023] | 0.298 | 0.000 | [-0.05, 0.05] | 0.993 |
| | | | HbA1C | 12,540 | 5.9% | -0.010 | [-0.025, 0.005] | 0.175 | 0.019* | [0.004, 0.033] | 0.014 | -0.013 | [-0.028, 0.002] | 0.094 |
| | | | Insulin | 168 | 0.1% | -0.099 | [-0.247, 0.05] | 0.191 | 0.029 | [-0.12, 0.178] | 0.701 | -0.070 | [-0.218, 0.077] | 0.348 |
| | | | BMI | 4,905 | 2.3% | 0.021 | [-0.006, 0.048] | 0.128 | 0.018 | [-0.009, 0.044] | 0.200 | -0.012 | [-0.039, 0.015] | 0.375 |
| | | | WC | 4,681 | 2.2% | 0.017 | [-0.007, 0.041] | 0.164 | 0.029* | [0.005, 0.053] | 0.018 | -0.017 | [-0.041, 0.007] | 0.163 |
| | | | Fasting glucose | † | † | -0.113 | [-0.247, 0.02] | 0.096 | 0.160* | [0.029, 0.291] | 0.017 | -0.067 | [-0.199, 0.064] | 0.314 |
| | | | HbA1C | 1,428 | 0.7% | 0.009 | [-0.035, 0.053] | 0.684 | 0.009 | [-0.035, 0.053] | 0.684 | -0.015 | [-0.059, 0.029] | 0.500 |
| EAS | 5,198 | 2.4% | Insulin | † | † | 0.256 | [-0.308, 0.819] | 0.332 | 0.160 | [-0.442, 0.762] | 0.562 | -0.007 | [-0.617, 0.603] | 0.981 |
| | | | BMI | 2,571 | 1.2% | 0.025 | [-0.012, 0.062] | 0.187 | 0.033 | [-0.004, 0.071] | 0.081 | 0.008 | [-0.029, 0.046] | 0.664 |
| | | | WC | 2,559 | 1.2% | 0.004 | [-0.028, 0.035] | 0.816 | 0.021 | [-0.01, 0.052] | 0.183 | 0.009 | [-0.022, 0.04] | 0.572 |
| | | | Fasting glucose | † | † | 0.003 | [-0.221, 0.227] | 0.981 | 0.061 | [-0.158, 0.279] | 0.580 | -0.172 | [-0.392, 0.048] | 0.124 |
| | | | HbA1C | 879 | 0.4% | 0.004 | [-0.051, 0.058] | 0.894 | 0.048 | [-0.007, 0.102] | 0.088 | -0.036 | [-0.091, 0.019] | 0.202 |
| SAS | 2,785 | 1.3% | Insulin | † | † | 0.280 | [-0.66, 1.22] | 0.525 | -0.001 | [-0.783, 0.782] | 0.998 | -0.215 | [-0.894, 0.464] | 0.501 |
| | | | BMI | 731 | 0.3% | -0.022 | [-0.093, 0.049] | 0.546 | 0.039 | [-0.031, 0.11] | 0.275 | -0.002 | [-0.073, 0.068] | 0.946 |
| | | | WC | 739 | 0.3% | 0.002 | [-0.056, 0.059] | 0.957 | 0.044 | [-0.013, 0.101] | 0.127 | -0.006 | [-0.062, 0.051] | 0.849 |
| | | | Fasting glucose | † | † | -0.265 | [-0.668, 0.137] | 0.183 | -0.302 | [-0.661, 0.057] | 0.094 | 0.427* | [0.1, 0.754] | 0.013 |
| | | | HbA1C | 321 | 0.2% | -0.060 | [-0.152, 0.033] | 0.209 | 0.052 | [-0.04, 0.145] | 0.267 | -0.016 | [-0.108, 0.076] | 0.736 |
| MID | 830 | 0.4% | Insulin | † | † | -0.365 | [-1.525, 0.795] | 0.309 | 0.459 | [-1.407, 2.325] | 0.401 | -0.304 | [-0.995, 0.386] | 0.199 |

Coefficients (β) from ordinary least squares (OLS) multivariable linear regression analysis of PRS for chronotype, short sleep, sleep duration on BMI, WC, fasting glucose, insulin, and hemoglobin A1C (HbA1C), adjusted for age, ancestry, ethnicity, sex, smoking, obesity, and diabetes. PRS and all measurements were standardized to a mean of 0 and a standard deviation of 1. Percentages are of the total 212,529 participants. Significance levels: *p < 0.05, **p < 0.01, ***p < 0.001. † Counts suppressed per the All of Us Data and Statistics Dissemination Policy.

## eTable 4. Longitudinal analyses results by ancestry

| | | | | | | PRS-C | | | PRS-SS | | | PRS-SD | | |
|---|---|---|---|---|---|---|---|---|---|---|---|---|---|---|
| **Genetic Ancestry** | **N** | **%** | **Measurement** | **N** | **%** | **β** | **95% CI** | **p-value** | **β** | **95% CI** | **p-value** | **β** | **95% CI** | **p-value** |
| | | | BMI | 74,094 | 34.9% | 0.002 | [-0.007, 0.011] | 0.637 | 0.017*** | [0.008, 0.025] | <0.001 | -0.016*** | [-0.026, -0.007] | <0.001 |
| | | | WC | 0 | 0.0% | N/A | N/A | N/A | N/A | N/A | N/A | N/A | N/A | N/A |
| | | | Fasting glucose | 1,728 | 0.8% | 0.009 | [-0.023, 0.041] | 0.573 | -0.016 | [-0.044, 0.012] | 0.262 | 0.028 | [-0.004, 0.061] | 0.084 |
| | | | HbA1C | 39,233 | 18.5% | -0.003 | [-0.011, 0.005] | 0.469 | 0.003 | [-0.005, 0.011] | 0.430 | -0.005 | [-0.013, 0.004] | 0.290 |
| All | 92,860 | 43.7% | Insulin | 135 | 0.1% | 0.091 | [-0.042, 0.224] | 0.181 | 0.134* | [0.008, 0.259] | 0.037 | -0.007 | [-0.109, 0.095] | 0.890 |
| | | | BMI | 50,573 | 23.8% | 0.003 | [-0.008, 0.013] | 0.607 | 0.017** | [0.006, 0.027] | 0.002 | -0.014** | [-0.024, -0.003] | 0.010 |
| | | | WC | 0 | 0.0% | N/A | N/A | N/A | N/A | N/A | N/A | N/A | N/A | N/A |
| | | | Fasting glucose | 910 | 0.4% | 0.020 | [-0.023, 0.063] | 0.368 | 0.013 | [-0.03, 0.056] | 0.549 | 0.029 | [-0.014, 0.073] | 0.190 |
| | | | HbA1C | 22,387 | 10.5% | 0.003 | [-0.007, 0.012] | 0.576 | 0.008 | [-0.002, 0.017] | 0.122 | -0.005 | [-0.015, 0.004] | 0.279 |
| EUR | 60,723 | 28.6% | Insulin | 63 | 0.0% | 0.081 | [-0.096, 0.258] | 0.372 | 0.229** | [0.072, 0.386] | 0.004 | -0.125* | [-0.247, -0.003] | 0.045 |
| | | | BMI | 11,314 | 5.3% | -0.011 | [-0.096, 0.258] | 0.331 | 0.009 | [-0.013, 0.031] | 0.421 | -0.010 | [-0.032, 0.012] | 0.379 |
| | | | WC | 0 | 0.0% | N/A | N/A | N/A | N/A | N/A | N/A | N/A | N/A | N/A |
| | | | Fasting glucose | 467 | 0.2% | 0.026 | [-0.031, 0.082] | 0.373 | -0.043 | [-0.098, 0.012] | 0.128 | 0.043 | [-0.013, 0.098] | 0.132 |
| AFR | 16,505 | 7.8% | HbA1C | 9,261 | 4.4% | -0.001 | [-0.018, 0.015] | 0.869 | -0.018* | [-0.035, -0.002] | 0.030 | 0.011 | [-0.005, 0.027] | 0.178 |
| | | | Insulin | 39 | 0.0% | 0.059 | [-0.085, 0.203] | 0.419 | 0.104 | [-0.076, 0.284] | 0.259 | -0.003 | N/A | N/A |
| | | | BMI | 9,212 | 4.3% | 0.018 | [-0.007, 0.043] | 0.156 | 0.021 | [-0.004, 0.046] | 0.100 | -0.030* | [-0.056, -0.004] | 0.024 |
| | | | WC | 0 | 0.0% | N/A | N/A | N/A | N/A | N/A | N/A | N/A | N/A | N/A |
| | | | Fasting glucose | 302 | 0.1% | -0.004 | [-0.083, 0.075] | 0.922 | -0.055 | [-0.136, 0.026] | 0.183 | 0.031 | [-0.049, 0.111] | 0.453 |
| | | | HbA1C | 6,274 | 3.0% | -0.014 | [-0.034, 0.006] | 0.158 | 0.013 | [-0.007, 0.033] | 0.192 | -0.019 | [-0.04, 0.001] | 0.060 |
| AMR | 12,176 | 5.7% | Insulin | 22 | 0.0% | 0.376 | [-0.067, 0.818] | 0.096 | 0.085 | [-0.441, 0.61] | 0.752 | 0.205 | [-0.115, 0.526] | 0.209 |
| | | | BMI | 1,745 | 0.8% | -0.007 | [-0.06, 0.046] | 0.793 | -0.018 | [-0.069, 0.033] | 0.497 | -0.033 | [-0.083, 0.017] | 0.196 |
| | | | WC | 0 | 0.0% | N/A | N/A | N/A | N/A | N/A | N/A | N/A | N/A | N/A |
| | | | Fasting glucose | † | † | -0.076 | [-0.167, 0.016] | 0.105 | 0.194 | [-0.145, 0.534] | 0.262 | 0.167** | [0.055, 0.278] | 0.003 |
| | | | HbA1C | 700 | 0.3% | -0.040 | [-0.094, 0.013] | 0.139 | 0.012 | [-0.041, 0.066] | 0.651 | 0.006 | [-0.047, 0.059] | 0.833 |
| EAS | 1,970 | 0.9% | Insulin | † | † | 0.087 | [-8.596, 8.77] | 0.984 | -0.482 | N/A | N/A | 0.683 | [-2.815, 4.182] | 0.702 |
| | | | BMI | 952 | 0.4% | 0.023 | [-0.05, 0.095] | 0.540 | 0.038 | [-0.033, 0.11] | 0.292 | -0.015 | [-0.086, 0.056] | 0.680 |
| | | | WC | 0 | 0.0% | N/A | N/A | N/A | N/A | N/A | N/A | N/A | N/A | N/A |
| | | | Fasting glucose | † | † | 0.069 | [-0.161, 0.298] | 0.558 | -0.190 | [-0.5, 0.12] | 0.230 | 0.139 | [-0.335, 0.612] | 0.566 |
| | | | HbA1C | 440 | 0.2% | -0.030 | [-0.11, 0.049] | 0.456 | 0.063 | [-0.018, 0.145] | 0.128 | 0.007 | [-0.074, 0.088] | 0.867 |
| SAS | 1,117 | 0.5% | Insulin | † | † | -0.184 | N/A | N/A | -0.337 | [-1.505, 0.83] | 0.571 | 0.193 | [-15.682, 16.068] | 0.981 |
| | | | BMI | 298 | 0.1% | -0.013 | [-0.129, 0.103] | 0.829 | -0.004 | [-0.12, 0.111] | 0.941 | 0.043 | [-0.075, 0.16] | 0.478 |
| | | | WC | 0 | 0.0% | N/A | N/A | N/A | N/A | N/A | N/A | N/A | N/A | N/A |
| MID | 369 | 0.2% | Fasting glucose | † | † | 0.512 | N/A | N/A | 0.471 | [-0.261, 1.202] | 0.207 | -0.018 | [-23.442, 23.407] | 0.999 |

| | | | HbA1C | † | † | -0.115 | [-0.242, 0.012] | 0.076 | 0.081 | [-0.031, 0.193] | 0.157 | 0.013 | [-0.108, 0.133] | 0.837 |
|---|---|---|---|---|---|---|---|---|---|---|---|---|---|---|
| | | | Insulin | 0 | 0.0% | N/A | N/A | N/A | N/A | N/A | N/A | N/A | N/A | N/A |

Fixed effects coefficients (β) from longitudinal analyses of PRS for sleep traits on BMI, WC, fasting glucose, hemoglobin A1C (HbA1C), insulin. PRS and all measurements were standardized to a mean of 0 and a standard deviation of 1. The linear mixed models were adjusted for ancestry, ethnicity, sex, smoking, obesity, and diabetes. Percentages are of the total 212,529 participants. Confidence intervals and p-values were unable to be computed for some parameters due to insufficient sample size or data variability. Significance levels: *p < 0.05, **p < 0.01, ***p < 0.001. † Counts suppressed per the All of Us Data and Statistics Dissemination Policy.

## eTable 5. Cross-sectional analyses results by ancestry, adjusted for Fitbit-measured sleep duration

| | | | | | | **PRS-C** | | | **Fitbit Sleep Duration (PRS-C)‡** | | | **PRS-SS** | | | **Fitbit Sleep Duration (PRS-SS)‡** | | | **PRS-SD** | | | **Fitbit Sleep Duration (PRS-SD)‡** | | |
|---|---|---|---|---|---|---|---|---|---|---|---|---|---|---|---|---|---|---|---|---|---|---|---|
| **Genetic Ancestry** | **N** | **%** | **Measurement** | **N** | **%** | **β** | **95% CI** | **p-value** | **β** | **95% CI** | **p-value** | **β** | **95% CI** | **p-value** | **β** | **95% CI** | **p-value** | **β** | **95% CI** | **p-value** | **β** | **95% CI** | **p-value** |
| | | | BMI | 7,091 | 3.3% | -0.006 | [-0.029, 0.017] | 0.599 | -0.090*** | [-0.112, -0.067] | 0.000 | 0.012 | [-0.01, 0.034] | 0.281 | -0.089*** | [-0.112, -0.067] | 0.000 | -0.018 | [-0.041, 0.005] | 0.127 | -0.089*** | [-0.111, -0.066] | 0.000 |
| | | | WC | 7,169 | 3.4% | 0.003 | [-0.017, 0.023] | 0.760 | -0.057*** | [-0.077, -0.037] | 0.000 | 0.005 | [-0.014, 0.025] | 0.587 | -0.057*** | [-0.076, -0.037] | 0.000 | -0.001 | [-0.021, 0.019] | 0.936 | -0.057*** | [-0.077, -0.037] | 0.000 |
| | | | Fasting glucose | 340 | 0.2% | -0.041 | [-0.141, 0.058] | 0.415 | -0.061 | [-0.156, 0.034] | 0.204 | -0.037 | [-0.131, 0.057] | 0.437 | -0.066 | [-0.161, 0.028] | 0.170 | 0.031 | [-0.066, 0.128] | 0.531 | -0.064 | [-0.159, 0.03] | 0.183 |
| | | | HbA1C | 2,558 | 1.2% | 0.010 | [-0.025, 0.046] | 0.568 | -0.028 | [-0.063, 0.007] | 0.118 | -0.027 | [-0.061, 0.007] | 0.119 | -0.030 | [-0.065, 0.005] | 0.098 | 0.018 | [-0.018, 0.054] | 0.316 | -0.029 | [-0.064, 0.006] | 0.103 |
| All | 7,655 | 3.6% | Insulin | 61 | <0.1% | 0.085 | [-0.241, 0.41] | 0.603 | 0.319* | [0.051, 0.587] | 0.021 | 0.089 | [-0.201, 0.379] | 0.541 | 0.309* | [0.042, 0.577] | 0.024 | 0.137 | [-0.161, 0.436] | 0.360 | 0.334* | [0.064, 0.603] | 0.016 |
| | | | BMI | 5,994 | 2.8% | -0.006 | [-0.03, 0.019] | 0.653 | -0.091*** | [-0.116, -0.067] | 0.000 | 0.018 | [-0.007, 0.042] | 0.157 | -0.091*** | [-0.115, -0.066] | 0.000 | -0.017 | [-0.041, 0.007] | 0.167 | -0.090*** | [-0.115, -0.066] | 0.000 |
| | | | WC | 6,075 | 2.9% | 0.000 | [-0.022, 0.021] | 0.967 | -0.057*** | [-0.078, -0.035] | 0.000 | 0.010 | [-0.011, 0.031] | 0.343 | -0.056*** | [-0.078, -0.035] | 0.000 | -0.001 | [-0.022, 0.021] | 0.954 | -0.057*** | [-0.078, -0.035] | 0.000 |
| | | | Fasting glucose | 287 | 0.1% | -0.044 | [-0.146, 0.059] | 0.401 | -0.082 | [-0.185, 0.021] | 0.117 | -0.040 | [-0.141, 0.062] | 0.443 | -0.088 | [-0.19, 0.015] | 0.093 | 0.054 | [-0.049, 0.156] | 0.302 | -0.085 | [-0.187, 0.018] | 0.104 |
| | | | HbA1C | 2,166 | 1.0% | 0.016 | [-0.021, 0.053] | 0.386 | -0.036 | [-0.073, 0.002] | 0.063 | -0.023 | [-0.06, 0.014] | 0.216 | -0.037* | [-0.075, -0.0] | 0.050 | 0.009 | [-0.028, 0.046] | 0.644 | -0.037 | [-0.074, 0.001] | 0.055 |
| EUR | 6,494 | 3.1% | Insulin | 43 | <0.1% | 0.131 | [-0.22, 0.482] | 0.452 | 0.282 | [-0.041, 0.606] | 0.085 | 0.033 | [-0.321, 0.388] | 0.849 | 0.271 | [-0.056, 0.599] | 0.101 | 0.132 | [-0.203, 0.468] | 0.428 | 0.305 | [-0.027, 0.636] | 0.071 |
| | | | BMI | 397 | 0.2% | -0.006 | [-0.102, 0.09] | 0.901 | -0.020 | [-0.118, 0.078] | 0.695 | -0.036 | [-0.134, 0.063] | 0.473 | -0.023 | [-0.121, 0.076] | 0.652 | -0.018 | [-0.115, 0.079] | 0.717 | -0.018 | [-0.116, 0.081] | 0.721 |
| | | | WC | 409 | 0.2% | 0.041 | [-0.045, 0.127] | 0.348 | -0.002 | [-0.089, 0.086] | 0.970 | -0.062 | [-0.149, 0.025] | 0.161 | -0.004 | [-0.092, 0.083] | 0.922 | 0.018 | [-0.069, 0.104] | 0.689 | -0.001 | [-0.088, 0.087] | 0.990 |
| | | | Fasting glucose | † | † | 0.063 | [-0.321, 0.447] | 0.734 | 0.062 | [-0.347, 0.471] | 0.755 | -0.023 | [-0.541, 0.496] | 0.927 | 0.065 | [-0.35, 0.48] | 0.744 | -0.302 | [-0.768, 0.164] | 0.189 | 0.127 | [-0.271, 0.525] | 0.509 |
| AFR | 426 | 0.2% | HbA1C | 180 | 0.1% | -0.042 | [-0.175, | 0.532 | 0.052 | [-0.083, | 0.445 | -0.052 | [-0.186, | 0.446 | 0.044 | [-0.091, | 0.520 | -0.009 | [-0.144, | 0.895 | 0.051 | [-0.086, | 0.465 |
| | | | | | | | 0.091] | | | 0.187] | | | 0.082] | | | 0.179] | | | 0.126] | | | 0.187] | |
| | | | Insulin | † | † | -0.345 | [-1.855, 1.164] | 0.560 | 0.915 | [-0.192, 2.023] | 0.083 | 0.327 | [-0.366, 1.021] | 0.279 | 0.603 | [-0.032, 1.238] | 0.059 | -0.044 | [-0.986, 0.897] | 0.903 | 0.794 | [-0.223, 1.812] | 0.096 |
| | | | BMI | 412 | 0.2% | 0.004 | [-0.089, 0.097] | 0.934 | -0.153** | [-0.246, -0.061] | 0.001 | -0.008 | [-0.1, 0.085] | 0.871 | -0.154** | [-0.247, -0.061] | 0.001 | -0.020 | [-0.114, 0.073] | 0.669 | -0.150** | [-0.244, -0.057] | 0.002 |
| | | | WC | 409 | 0.2% | 0.015 | [-0.07, 0.099] | 0.733 | -0.111* | [-0.195, -0.026] | 0.011 | -0.003 | [-0.087, 0.081] | 0.937 | -0.110* | [-0.195, -0.025] | 0.011 | -0.030 | [-0.115, 0.054] | 0.481 | -0.106* | [-0.191, -0.02] | 0.016 |
| | | | Fasting glucose | † | † | -0.539 | [-1.227, 0.149] | 0.112 | -0.083 | [-0.671, 0.505] | 0.760 | -0.033 | [-0.597, 0.531] | 0.899 | -0.096 | [-0.777, 0.585] | 0.760 | -0.174 | [-0.812, 0.463] | 0.556 | -0.047 | [-0.724, 0.631] | 0.881 |
| | | | HbA1C | 134 | 0.1% | 0.013 | [-0.146, 0.172] | 0.876 | -0.005 | [-0.16, 0.149] | 0.946 | -0.051 | [-0.206, 0.105] | 0.521 | -0.006 | [-0.16, 0.148] | 0.937 | 0.106 | [-0.049, 0.261] | 0.177 | -0.025 | [-0.18, 0.13] | 0.749 |
| AMR | 435 | 0.2% | Insulin | † | † | 0.808 | N/A | N/A | 0.616 | N/A | N/A | 0.273 | N/A | N/A | 1.154 | N/A | N/A | 0.793 | [-3.181, 4.768] | 0.239 | 0.434 | [-3.54, 4.408] | 0.398 |
| | | | BMI | 168 | 0.1% | -0.011 | [-0.16, 0.138] | 0.887 | -0.151* | [-0.301, -0.001] | 0.048 | 0.050 | [-0.099, 0.198] | 0.511 | -0.149 | [-0.299, 0.001] | 0.052 | -0.114 | [-0.264, 0.036] | 0.134 | -0.139 | [-0.289, 0.011] | 0.070 |
| EAS | 176 | 0.1% | WC | 161 | 0.1% | 0.039 | [-0.095, 0.172] | 0.569 | -0.217** | [-0.35, -0.085] | 0.001 | 0.022 | [-0.111, 0.155] | 0.742 | -0.217** | [-0.35, -0.085] | 0.001 | -0.097 | [-0.231, 0.038] | 0.158 | -0.203* | [-0.337, -0.07] | 0.003 |
| | | | Fasting glucose | † | † | 0.376 | [-3.921, 4.674] | 0.466 | 0.325 | [-3.47, 4.119] | 0.473 | 0.146 | [-1.062, 1.355] | 0.655 | 0.228 | [-0.98, 1.436] | 0.502 | -0.586 | [-5.512, 4.339] | 0.372 | 0.311 | [-2.752, 3.375] | 0.419 |
| | | | HbA1C | † | † | -0.045 | [-0.305, 0.214] | 0.726 | -0.013 | [-0.278, 0.253] | 0.923 | 0.149 | [-0.105, 0.402] | 0.243 | -0.023 | [-0.278, 0.232] | 0.857 | 0.148 | [-0.14, 0.436] | 0.305 | -0.071 | [-0.343, 0.202] | 0.602 |
| | | | Insulin | † | † | 0.000 | N/A | N/A | 0.000 | N/A | N/A | 0.000 | N/A | N/A | 0.000 | N/A | N/A | 0.000 | N/A | N/A | 0.000 | N/A | N/A |
| | | | BMI | 106 | <0.1 % | -0.079 | [-0.267, 0.109] | 0.406 | -0.194 | [-0.39, 0.001] | 0.051 | -0.101 | [-0.291, 0.089] | 0.293 | -0.186 | [-0.38, 0.008] | 0.060 | 0.175 | [-0.008, 0.358] | 0.060 | -0.183 | [-0.374, 0.009] | 0.062 |
| | | | WC | 102 | <0.1 % | -0.081 | [-0.254, 0.092] | 0.355 | -0.011 | [-0.19, 0.168] | 0.901 | -0.124 | [-0.299, 0.05] | 0.161 | -0.009 | [-0.187, 0.169] | 0.920 | 0.186* | [0.021, 0.351] | 0.028 | -0.010 | [-0.185, 0.165] | 0.910 |
| | | | Fasting glucose | † | † | 0.500 | N/A | N/A | -0.500 | N/A | N/A | 0.500 | N/A | N/A | -0.500 | N/A | N/A | 0.500 | N/A | N/A | -0.500 | N/A | N/A |
| | | | HbA1C | † | † | 0.045 | [-0.39, 0.48] | 0.832 | -0.132 | [-0.597, 0.334] | 0.563 | -0.034 | [-0.522, 0.453] | 0.886 | -0.123 | [-0.589, 0.342] | 0.588 | 0.317 | [-0.101, 0.735] | 0.130 | -0.090 | [-0.531, 0.351] | 0.676 |
| SAS | † | † | Insulin | † | † | 0.000 | N/A | N/A | 0.000 | N/A | N/A | 0.000 | N/A | N/A | 0.000 | N/A | N/A | 0.000 | N/A | N/A | 0.000 | N/A | N/A |
| | | | BMI | † | † | 0.076 | [-0.59, 0.741] | 0.796 | 0.067 | [-0.498, 0.631] | 0.788 | 0.080 | [-0.476, 0.636] | 0.744 | 0.082 | [-0.474, 0.637] | 0.739 | -0.169 | [-0.699, 0.362] | 0.477 | 0.059 | [-0.482, 0.6] | 0.803 |
| | | | WC | † | † | 0.234 | [-0.567, 1.034] | 0.487 | -0.415 | [-1.16, 0.33] | 0.212 | 0.008 | [-0.952, 0.967] | 0.985 | -0.404 | [-1.23, 0.423] | 0.265 | 0.164 | [-0.572, 0.9] | 0.592 | -0.419 | [-1.183, 0.344] | 0.217 |
| MID | † | † | Fasting glucose | † | † | 0.000 | N/A | N/A | 0.000 | N/A | N/A | 0.000 | N/A | N/A | 0.000 | N/A | N/A | 0.000 | N/A | N/A | 0.000 | N/A | N/A |

| Genetic Ancestry | N | % | Measurement | N | % | PRS-C β | PRS-C 95% CI | PRS-C p-value | Fitbit Sleep Duration (PRS-C)‡ β | 95% CI | p-value | PRS-SS β | PRS-SS 95% CI | PRS-SS p-value | Fitbit Sleep Duration (PRS-SS)‡ β | 95% CI | p-value | PRS-SD β | PRS-SD 95% CI | PRS-SD p-value | Fitbit Sleep Duration (PRS-SD)‡ β | 95% CI | p-value |
|---|---|---|---|---|---|---|---|---|---|---|---|---|---|---|---|---|---|---|---|---|---|---|---|
| | | | HbA1C | † | † | 0.668 | N/A | N/A | 1.277 | N/A | N/A | 0.368 | N/A | N/A | 1.117 | N/A | N/A | -0.312 | N/A | N/A | 1.064 | N/A | N/A |
| | | | Insulin | 0 | 0.0% | N/A | N/A | N/A | N/A | N/A | N/A | N/A | N/A | N/A | N/A | N/A | N/A | N/A | N/A | N/A | N/A | N/A | N/A |

Coefficients (β) from ordinary least squares (OLS) multivariable linear regression analysis of PRS for chronotype, short sleep, sleep duration on BMI, WC, fasting glucose, insulin, and hemoglobin A1C (HbA1c), adjusted for Fitbit-derived sleep duration, age, ancestry, ethnicity, sex, smoking, obesity, and diabetes. PRS and all measurements were standardized to a mean of 0 and a standard deviation of 1. Percentages are of the total 212,529 participants. Significance levels: *p < 0.05, **p < 0.01, ***p < 0.001. † Counts suppressed per the All of Us Data and Statistics Dissemination Policy. ‡ Association test between Fitbit-derived sleep duration and health outcomes, with PRS included as a covariate..

## eTable 6. Longitudinal analyses results by ancestry, adjusted for Fitbit-measured sleep duration

| | | | | | | **PRS-C** | | | **Fitbit Sleep Duration (PRS-C)‡** | | | **PRS-SS** | | | **Fitbit Sleep Duration (PRS-SS)‡** | | | **PRS-SD** | | | **Fitbit Sleep Duration (PRS-SD)‡** | | |
|---|---|---|---|---|---|---|---|---|---|---|---|---|---|---|---|---|---|---|---|---|---|---|---|
| **Genetic Ancestry** | **N** | **%** | **Measurement** | **N** | **%** | **β** | **95% CI** | **p-value** | **β** | **95% CI** | **p-value** | **β** | **95% CI** | **p-value** | **β** | **95% CI** | **p-value** | **β** | **95% CI** | **p-value** | **β** | **95% CI** | **p-value** |
| | | | BMI | 3,641 | 1.7% | 0.006 | [-0.036, 0.047] | 0.790 | -0.024 | [-0.065, 0.018] | 0.261 | 0.025 | [-0.015, 0.065] | 0.225 | -0.023 | [-0.065, 0.018] | 0.268 | -0.041 | [-0.082, 0.001] | 0.053 | -0.022 | [-0.063, 0.02] | 0.308 |
| | | | WC | 0 | 0.0% | N/A | N/A | N/A | N/A | N/A | N/A | N/A | N/A | N/A | N/A | N/A | N/A | N/A | N/A | N/A | N/A | N/A | N/A |
| | | | Fasting glucose | † | † | -0.138 | [-0.329, 0.052] | 0.155 | -0.055 | [-0.213, 0.103] | 0.493 | 0.008 | [-0.136, 0.152] | 0.913 | -0.077 | [-0.239, 0.086] | 0.356 | 0.053 | [-0.134, 0.24] | 0.578 | -0.067 | [-0.235, 0.101] | 0.435 |
| | | | Hemoglobin A1C | 1,318 | 0.6% | -0.006 | [-0.044, 0.032] | 0.754 | -0.035 | [-0.073, 0.002] | 0.066 | -0.009 | [-0.045, 0.027] | 0.625 | -0.036 | [-0.074, 0.002] | 0.064 | -0.004 | [-0.042, 0.035] | 0.851 | -0.035 | [-0.073, 0.003] | 0.070 |
| All | 4,078 | 1.9% | Insulin | † | † | 0.287 | [-0.375, 0.948] | 0.396 | -0.042 | [-0.836, 0.753] | 0.918 | 0.140 | N/A | N/A | -0.039 | N/A | N/A | -0.082 | [-0.411, 0.247] | 0.626 | 0.074 | [-0.365, 0.512] | 0.742 |
| EUR | 3,560 | 1.7% | BMI | 3,198 | 1.5% | 0.007 | [-0.037, | 0.767 | -0.026 | [-0.071, | 0.253 | 0.028 | [-0.015, | 0.203 | -0.026 | [-0.07, | 0.259 | -0.030 | [-0.074, | 0.177 | -0.025 | [-0.069, | 0.276 |
| | | | | | | | 0.05] | | | 0.019] | | | 0.071] | | | 0.019] | | | 0.014] | | | 0.02] | |
| | | | WC | 0 | 0.0% | N/A | N/A | N/A | N/A | N/A | N/A | N/A | N/A | N/A | N/A | N/A | N/A | N/A | N/A | N/A | N/A | N/A | N/A |
| | | | Fasting glucose | † | † | -0.156 | [-0.347, 0.035] | 0.110 | -0.053 | [-0.195, 0.088] | 0.459 | 0.001 | [-0.158, 0.161] | 0.988 | -0.086 | [-0.27, 0.097] | 0.358 | 0.065 | [-0.177, 0.308] | 0.597 | -0.068 | [-0.27, 0.133] | 0.506 |
| | | | Hemoglobin A1C | 1,097 | 0.5% | 0.000 | [-0.041, 0.04] | 0.993 | -0.044* | [-0.086, -0.002] | 0.041 | -0.003 | [-0.043, 0.038] | 0.891 | -0.044* | [-0.086, -0.002] | 0.040 | -0.010 | [-0.051, 0.031] | 0.619 | -0.043* | [-0.085, -0.001] | 0.043 |
| | | | Insulin | † | † | 0.250 | [-9.636, 10.135] | 0.961 | -0.066 | [-3.727, 3.596] | 0.972 | 0.086 | N/A | N/A | 0.071 | N/A | N/A | -0.115 | [-4.712, 4.482] | 0.961 | 0.049 | [-0.851, 0.95] | 0.914 |
| | | | BMI | 170 | 0.1% | -0.036 | [-0.222, 0.15] | 0.706 | 0.087 | [-0.093, 0.267] | 0.341 | -0.129 | [-0.319, 0.061] | 0.182 | 0.080 | [-0.097, 0.257] | 0.377 | -0.119 | [-0.291, 0.053] | 0.176 | 0.105 | [-0.075, 0.285] | 0.254 |
| | | | WC | 0 | 0.0% | N/A | N/A | N/A | N/A | N/A | N/A | N/A | N/A | N/A | N/A | N/A | N/A | N/A | N/A | N/A | N/A | N/A | N/A |
| | | | Fasting glucose | † | † | -0.932* | [-1.827, -0.036] | 0.041 | 1.431 | [-0.158, 3.021] | 0.078 | -0.546 | [-2.24, 1.147] | 0.527 | 2.243 | [-1.01, 5.495] | 0.177 | -0.794 | [-2.028, 0.44] | 0.207 | 2.353** | [0.717, 3.989] | 0.005 |
| | | | Hemoglobin A1C | † | † | -0.009 | [-0.119, 0.101] | 0.867 | -0.019 | [-0.132, 0.093] | 0.737 | -0.040 | [-0.147, 0.066] | 0.457 | -0.028 | [-0.141, 0.085] | 0.631 | 0.020 | [-0.077, 0.117] | 0.692 | -0.023 | [-0.136, 0.089] | 0.683 |
| AFR | 215 | 0.1% | Insulin | † | † | N/A | N/A | N/A | N/A | N/A | N/A | N/A | N/A | N/A | N/A | N/A | N/A | N/A | N/A | N/A | N/A | N/A | N/A |
| | | | BMI | 165 | 0.1% | -0.034 | [-0.218, 0.15] | 0.719 | -0.097 | [-0.267, 0.074] | 0.266 | 0.001 | [-0.183, 0.185] | 0.991 | -0.097 | [-0.267, 0.074] | 0.267 | -0.071 | [-0.246, 0.104] | 0.427 | -0.088 | [-0.259, 0.084] | 0.318 |
| AMR | 185 | 0.1% | WC | 0 | 0.0% | N/A | N/A | N/A | N/A | N/A | N/A | N/A | N/A | N/A | N/A | N/A | N/A | N/A | N/A | N/A | N/A | N/A | N/A |
| | | | Fasting glucose | † | † | N/A | N/A | N/A | N/A | N/A | N/A | N/A | N/A | N/A | N/A | N/A | N/A | N/A | N/A | N/A | N/A | N/A | N/A |
| | | | Hemoglobin A1C | † | † | -0.022 | [-0.199, 0.155] | 0.808 | 0.065 | [-0.107, 0.236] | 0.459 | 0.006 | [-0.141, 0.153] | 0.937 | 0.070 | [-0.096, 0.237] | 0.408 | 0.081 | [-0.101, 0.264] | 0.381 | 0.061 | [-0.107, 0.229] | 0.477 |
| | | | Insulin | † | † | N/A | N/A | N/A | N/A | N/A | N/A | N/A | N/A | N/A | N/A | N/A | N/A | N/A | N/A | N/A | N/A | N/A | N/A |
| | | | BMI | 69 | <0.1 % | 0.253 | [-0.097, 0.603] | 0.156 | 0.134 | [-0.298, 0.566] | 0.543 | 0.196 | [-0.157, 0.549] | 0.277 | 0.158 | [-0.277, 0.594] | 0.476 | -0.102 | [-0.436, 0.233] | 0.551 | 0.164 | [-0.276, 0.605] | 0.465 |
| | | | WC | 0 | 0.0% | N/A | N/A | N/A | N/A | N/A | N/A | N/A | N/A | N/A | N/A | N/A | N/A | N/A | N/A | N/A | N/A | N/A | N/A |
| | | | Fasting glucose | † | † | N/A | N/A | N/A | N/A | N/A | N/A | N/A | N/A | N/A | N/A | N/A | N/A | N/A | N/A | N/A | N/A | N/A | N/A |
| | | | Hemoglobin A1C | † | † | 0.093 | [-0.154, 0.34] | 0.462 | -0.088 | [-0.457, 0.28] | 0.639 | 0.048 | [-0.216, 0.312] | 0.721 | -0.110 | [-0.479, 0.259] | 0.559 | 0.604** | [0.178, 1.031] | 0.006 | -0.542** | [-0.95, -0.134] | 0.009 |
| EAS | 75 | <0.1 % | Insulin | 0 | 0.0% | N/A | N/A | N/A | N/A | N/A | N/A | N/A | N/A | N/A | N/A | N/A | N/A | N/A | N/A | N/A | N/A | N/A | N/A |
| | | | BMI | † | † | 0.038 | [-0.295, 0.372] | 0.822 | -0.262 | [-0.646, 0.121] | 0.180 | 0.226 | [-0.11, 0.562] | 0.187 | -0.240 | [-0.59, 0.11] | 0.179 | -0.216 | [-0.516, 0.084] | 0.158 | -0.271 | [-0.637, 0.096] | 0.148 |
| | | | WC | 0 | 0.0% | N/A | N/A | N/A | N/A | N/A | N/A | N/A | N/A | N/A | N/A | N/A | N/A | N/A | N/A | N/A | N/A | N/A | N/A |
| | | | Fasting glucose | 0 | 0.0% | N/A | N/A | N/A | N/A | N/A | N/A | N/A | N/A | N/A | N/A | N/A | N/A | N/A | N/A | N/A | N/A | N/A | N/A |
| | | | Hemoglobin A1C | † | † | 0.319*** | [0.255, 0.383] | 0.000 | -0.051 | [-0.395, 0.292] | 0.770 | -0.499*** | [-0.761, -0.237] | 0.000 | 0.143 | [-0.096, 0.381] | 0.241 | 0.265 | N/A | N/A | -0.040 | [-0.376, 0.297] | 0.817 |
| SAS | † | † | Insulin | 0 | 0.0% | N/A | N/A | N/A | N/A | N/A | N/A | N/A | N/A | N/A | N/A | N/A | N/A | N/A | N/A | N/A | N/A | N/A | N/A |
| MID | † | † | BMI | † | † | 0.531 | [-0.61, | 0.362 | 1.092 | [-0.096, | 0.072 | 2.374 | [-3.207, | 0.404 | 3.103 | [-2.452, | 0.274 | -0.556 | [-1.975, | 0.443 | 1.234 | [-0.369, | 0.131 |

| | | | | | | | 1.67 2] | | | 2.27 9] | | | 7.95 5] | | | 8.65 7] | | | 0.86 3] | | | 2.83 7] | |
|---|---|---|---|---|---|---|---|---|---|---|---|---|---|---|---|---|---|---|---|---|---|---|---|
| | | | WC | 0 | 0.0% | N/A | N/A | N/A | N/A | N/A | N/A | N/A | N/A | N/A | N/A | N/A | N/A | N/A | N/A | N/A | N/A | N/A | N/A |
| | | | Fasting glucose | 0 | 0.0% | N/A | N/A | N/A | N/A | N/A | N/A | N/A | N/A | N/A | N/A | N/A | N/A | N/A | N/A | N/A | N/A | N/A | N/A |
| | | | Hemoglobin A1C | † | † | N/A | s | N/A | N/A | N/A | N/A | N/A | N/A | N/A | N/A | N/A | N/A | N/A | N/A | N/A | N/A | N/A | N/A |
| | | | Insulin | 0 | 0.0% | N/A | N/A | N/A | N/A | N/A | N/A | N/A | N/A | N/A | N/A | N/A | N/A | N/A | N/A | N/A | N/A | N/A | N/A |

Fixed effects coefficients (β) from longitudinal analyses of PRS for sleep traits on BMI, WC, fasting glucose, hemoglobin A1C, insulin. PRS and all measurements were standardized to a mean of 0 and a standard deviation of 1. The linear mixed models were adjusted for Fitbit-derived sleep duration, ancestry, ethnicity, sex, smoking, obesity, and diabetes. Percentages are of the total 212,529 participants. Confidence intervals and p-values were unable to be computed for some parameters due to insufficient sample size or data variability. Significance levels: *p < 0.05, **p < 0.01, ***p < 0.001. † Counts suppressed per the All of Us Data and Statistics Dissemination Policy. ‡ Association test between Fitbit-derived sleep duration and health outcomes, with PRS included as a covariate..

**eTable 7. PRS vs. sleep duration contribution analysis.** We quantified the relative contributions of sleep duration (SD) compared to PRS-SD and PRS-SS using coefficients (β) from a linear mixed model for associations that were significant without sleep duration as a covariate but became non-significant after its inclusion, as shown in Tables 2 and 3.

| | PRS-SS | | | PRS-SD | | |
|---|---|---|---|---|---|---|
| | $\beta_{PRS}$ | $\beta_{SD}$ | $Contribution_{SD}$ | $\beta_{PRS}$ | $\beta_{SD}$ | $Contribution_{SD}$ |
| **Cross-sectional analyses** | | | | | | |
| BMI | 0.012 | -0.089 | 98.2% | -0.018 | -0.089 | 96.1% |
| WC | 0.005 | -0.056 | 99.2% | -0.001 | -0.056 | 99.9% |
| Insulin | N/A | N/A | N/A | 0.137 | 0.334 | 85.6% |
| **Longitudinal analyses** | | | | | | |
| BMI | 0.024 | -0.021 | 44.0% | -0.041 | -0.020 | 18.7% |
| Insulin | 0.140 | -0.039 | 7.1% | N/A | N/A | N/A |

- Example formula used for linear mixed model: wc ~ PRS_SS + person_id + smoking + obesity + diabetes + sleep_duration + age_at_measurement + ancestry_pred_afr + ancestry_pred_amr + ancestry_pred_eas + ancestry_pred_mid + ancestry_pred_sas + ethnicity_Hispanic_or_Latino + sex_at_birth_Male
- $\text{Contribution}_{\text{sleep duration}} = \frac{{\beta_{SD}}^2}{{\beta_{PRS}}^2 + {\beta_{SD}}^2} \times 100\ [\%]$